\documentclass{aa}  

\usepackage{graphicx}
\usepackage{multicol}
\usepackage{txfonts}
\usepackage{hyperref}
\usepackage{xcolor}

\begin{document}

   \title{Turnaround radius of galaxy clusters in N-body simulations}


   \author{Giorgos Korkidis,
          \inst{1}\fnmsep\inst{2}\fnmsep\thanks{E-mail: gkorkidis@physics.uoc.gr},
          Vasiliki Pavlidou \inst{1}\fnmsep\inst{2}\fnmsep\thanks{E-mail: pavlidou@physics.uoc.gr},
          Konstantinos Tassis,\inst{1}\fnmsep\inst{2},
          Evangelia Ntormousi\inst{1}\fnmsep\inst{2}, 
          Theodore N. Tomaras\inst{1},
          Konstantinos Kovlakas\inst{1}\fnmsep\inst{2}
          }
    
    \authorrunning{G. Korkidis et al.}

   \institute{Department of Physics and Institute for Theoretical and Computational Physics, University of Crete, GR-70013 Heraklio, Greece
         \and
             Institute of Astrophysics, Foundation for Research and Technology – Hellas, Vassilika Vouton, GR-70013 Heraklio, Greece
             }

   \date{}

 
  \abstract
  {}
   {We use N-body simulations to examine whether a characteristic turnaround radius, as predicted from the spherical collapse model in a $\rm {\Lambda CDM}$ Universe, can be meaningfully identified for galaxy clusters, in the presence of full three-dimensional effects.}{We use The Dark Sky Simulations and Illustris-TNG dark-matter--only cosmological runs to calculate radial velocity profiles around collapsed structures, extending out to many times the virial radius $R_{200}$. There, the turnaround radius can be unambiguously identified as the largest non-expanding scale around a center of gravity.} {We find that: (a) Indeed, a single turnaround scale can meaningfully describe strongly non-spherical structures. (b) For halos of masses $M_{200}>10^{13}M_\odot$, the turnaround radius $R_{ta}$ scales with the enclosed mass $M_{ta}$ as $M_{ta}^{1/3}$, as predicted by the spherical collapse model. (c) The deviation of $R_{ta}$ in simulated halos from the spherical collapse model prediction is rather insensitive to halo asphericity. Rather, it is sensitive to the tidal forces due to massive neighbors when such are present.  (d) Halos exhibit a characteristic average density within the turnaround scale. This characteristic density is dependent on cosmology and redshift. For the present cosmic epoch and for concordance cosmological parameters ($\Omega_m \sim 0.7$; $\Omega_\Lambda \sim 0.3$) turnaround structures exhibit an average matter density contrast with the background Universe of $\delta \sim 11$. Thus $R_{ta}$  is equivalent to $R_{11}$ -- in a way analogous to defining the "virial" radius as $R_{200}$ -- with the advantage that $R_{11}$ is shown in this work to correspond to a kinematically relevant scale in N-body simulations }{}


   \maketitle
\section{Introduction} \label{section 1}
The turnaround radius naturally appears within the context of the collapse of a single, spherically symmetric structure in an otherwise homogeneous and isotropic expanding universe, as the boundary between the non-expanding structure and the Hubble flow.
In recent years the turnaround radius has attracted considerable attention (e.g., \citealp{Pavlidou_Tomaras},\citealp{Tanoglidis2015}, \citealp{LY16}, \citealp{BT17}, \citealp{BDRST17}, \citealp{Leesix}, \citealp{LVAS18}, \citealp{NOF18}, \citealp{LVA19}, \citealp{CDL19}) due to its potential as a possible new probe of cosmological parameters, as a constraint on alternative theories of gravity, and as a well-defined boundary for large-scale structures.

The attractiveness of the turnaround radius as a boundary for cosmic structures stems from two factors. The first one is its unambiguous definition, based on the radial velocity profile, which on the one hand speaks to the dynamics on the structure, and on the other hand is straight-forward to both calculate and explain: the turnaround radius is the scale where the edge of the structure joins the Hubble flow. The second is that  structures defined on turnaround scales have only mildly evolved into the non-linear regime. Thus their behavior is expected to be closer to the predictions of simple analytic models. 

The potential of the turnaround radius as a cosmological observable has been primarily explored under the assumption of a single, spherically symmetric structure evolving in an otherwise unperturbed Universe. The "spherical collapse" calculations that follow from this assumption make the following predictions (\citealp{Pavlidou_Tomaras, Tetal16}, Pavlidou et al. 2019 in prep):
\begin{itemize}
    \item[(a)] The matter enclosed by the turnaround radius has a characteristic average matter density (the "turnaround density", $\rho_{ta}$), which is the same for structures of all masses at a given cosmic epoch.
    \item[(b)] The value of $\rho_{ta}$ and its evolution with cosmic time depends on (and probes) the cosmological parameters $\Omega_m$ and $\Omega_\Lambda$.
    \item[(c)] At late cosmic times, when $\Lambda$ fully dominates the dynamics of the expansion, $\rho_{ta}$ asymptotically approaches a value which is only dependent on the value of the cosmological constant and is equal to $2\rho_\Lambda$, where $\rho_\Lambda=\Lambda c^2/8\pi G$.
    \item[(d)] As a consequence, the radius of any non-expanding structure of mass $M$ in a $\Lambda$CDM universe can never exceed $(3GM/\Lambda c^2)^{1/3}$.
\end{itemize} 

However, before any comparisons between these predictions of the spherical collapse model and observational data can be meaningfully made, it is necessary to test whether the predictions persist in  simulated structures with realistic shapes, and to quantify any deviations due to departures from spherical symmetry and the presence of neighboring structures.  This is the scope of the current paper. In  particular, we address the following questions:
\begin{enumerate}
    \item Can a single turnaround radius meaningfully characterize a realistic 3D structure?
    \item If so, how does this turnaround radius compare to the predictions of the spherical collapse model for objects of the same mass?
    \item Does the prediction of spherical collapse that a characteristic average turnaround density, $\rho_{ta}$, exists for all structures at a given redshift persist in N-body simulations?
    \item Does the turnaround radius depend on the shape of structures?
      \item Does the turnaround radius depend on the presence of massive neighbors?
\end{enumerate}

This paper is structured as follows. In \S \ref{section 2} we describe the set of N-body simulations used in this work and the characteristics of the dark matter halos under study. In \S \ref{section 3} we describe our methodology for the calculation of the turnaround radius and for substructure removal when the turnaround radius is used as the boundary of a structure. In \S \ref{section 4} we compare the turnaround radius in simulated structures to the predictions of the spherical collapse model and investigate the effect of halo shape and environment on the turnaround radius. We discuss these findings in section \S \ref{section discussion} and conclude in \S \ref{section 5}. 


\section{N-body simulations} \label{section 2}
In this work we use data from: (a) The Dark Sky Simulations \citep{darksky} (hereafter TDSS). TDSS were run using 2HOT \citep{2HOT} a purely adaptive tree-based N-body code. The data extraction and analysis was made using yt \citep{yt}. (b)  The Illustris-TNG simulations \citep{Illustris}, which is a suite of gravo-magnetohydrodynamical simulations run with the AREPO moving mesh code \citep{AREPO}. For the data extraction of the Illustris-TNG we used the methods presented on the project's website \footnote{http://www.tng-project.org/data/} along with yt. We list the basic properties of the simulations we use in \autoref{table 1}.  \autoref{table 2} shows the cosmological parameters adopted in each simulation. 

For producing halo catalogs, TDSS uses the ROCKSTAR algorithm \citep{ROCKSTAR} and Illustris-TNG uses friends-of-friends (FOF) \citep{FOF}. FOF is historically one of the first algorithms used to identify halos. It considers particles to be members of the same group if their distance is smaller than a given linking length \citep{Kravtsov,FOF}. The value of the linking length is usually chosen so that the defined halo has a density contrast that approaches that of a virialized structure \citep{Kravtsov, ROCKSTAR, HalosGoneMad}.  ROCKSTAR  makes use of the full six-dimensional phase space of particle positions and velocities, as well as of time \citep{ROCKSTAR}. 

The snapshot that we used from each simulation was  $z=0$ (the present cosmic epoch).

\begin{table}[htb!]
\begin{center}
\caption{Simulation  number of particles, box size, and particle mass. Both simulations are from dark-matter--only runs.}
\label{table 1}
\begin{tabular}{c c c c}
\hline
Simulation & $\sqrt[3]{N}$ & $L\left [ h^{-1}  Mpc \right ]$   & $M_p\left [ h^{-1} M_\odot  \right ]$  \\ \hline
ds14\_a       & $10240$ & $8000$ & $3.90 \ 10^{10}$  \\
TNG300-3-Dark & $625$ & $205$ & $3.02 \ 10^{9}$ \\ \hline
\end{tabular}
\end{center}
\end{table}

\begin{table}[htb!]
\centering
\caption{Cosmological parameters used in simulations}
\label{table 2}
\begin{tabular}{c c c c c}
\hline
Simulation & $\Omega_{m,0}$ & $\Omega_{b,0}$ & $\Omega_{{\Lambda,0}}$ & $h_{100}$ \\ \hline
ds14\_a  & $0.295$ & $0.0468$ & $0.705$  & $0.688$  \\
TNG300-3-Dark & $0.308$ & $0.0486$ & $0.691$  & $0.677$ \\ \hline
\end{tabular}
\end{table}

The resolution provided by both simulations is more than adequate in order to have a sufficient number of particles within turnaround scales. Some special consideration regarding the removal of substructure is necessary for our analysis. Although each halo finder does include a substructure identification algorithm \citep{Subhalos}, these operate on the virial scale (or a scale that is purposefully selected to be close to the virial scale, in the case of FOF halo finders). We however are interested in placing the boundary of structures at the much larger turnaround scale. As a consequence, there may be structures that are located outside the virial radii of all nearby larger structures (hence not considered "substructure" by the halo finding algorithms), but within some structure's turnaround radius (and are thus "substructure" for our purposes).

A somewhat similar problem arises for structures in the process of merging. If the centers of mass of two structures are approaching each other, then the two structures are a part of a single "turnaround" structure (they are not receding from each other as the universe expands, hence the system as a whole has detached from the Hubble flow). However, the distance between the two centers may still be large enough that the halo-finding algorithm lists these as two distinct structures. This may result in the same structure appearing twice in our analysis. The larger "turnaround" structure encompassing the system may also have a center of mass significantly displaced from the center of either individual structure provided by the halo-finding algorithm. We discuss our strategies for dealing with this issue in \S \ref{section 3}.

For computational efficiency, we only analyzed a (randomly chosen) fraction of the structures found in each simulation box. From TDSS we randomly chose 578 halos with\footnote{Smaller halos were not considered because they were not included in the halo catalog at the time of writing this paper.} $\rm M_{200} \geq 10^{15}\rm M_\odot$  while from the Illustris-TNG we chose 438 halos with masses in the interval $\rm{ 10^{13} M_\odot\leq M_{200} \leq 10^{15} M_\odot}$ and which were residing in a randomly chosen\footnote{The region was chosen close to the center of the box in order to avoid possible boundary effects.} spherical region of $R = 100 \rm{Mpc}$ radius. 

\section{Calculating the turnaround radius} \label{section 3}
   \begin{figure*}[htb!]
    \includegraphics[width=1.07\columnwidth,clip]{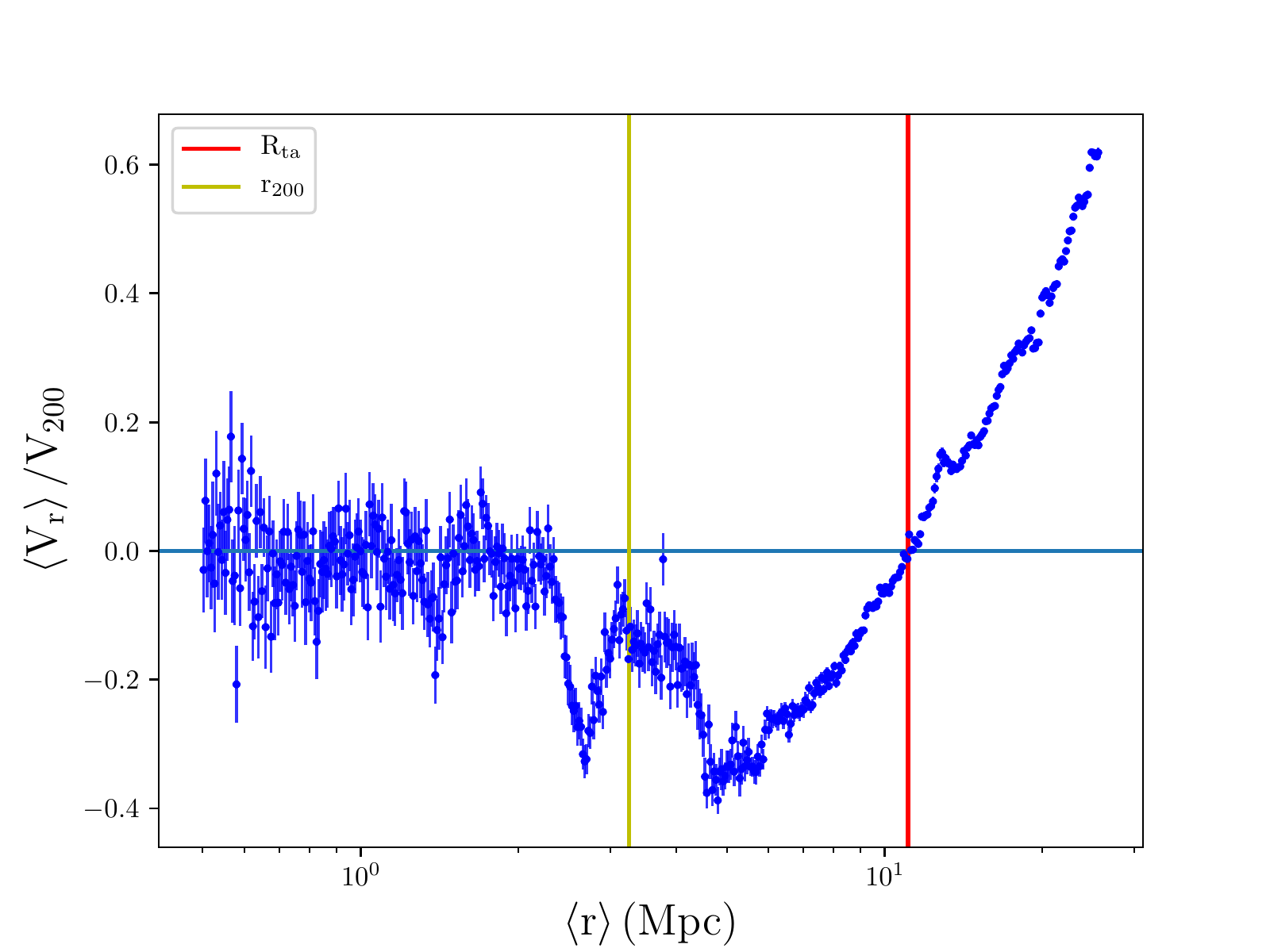}
      \includegraphics[width=1.07\columnwidth,clip]{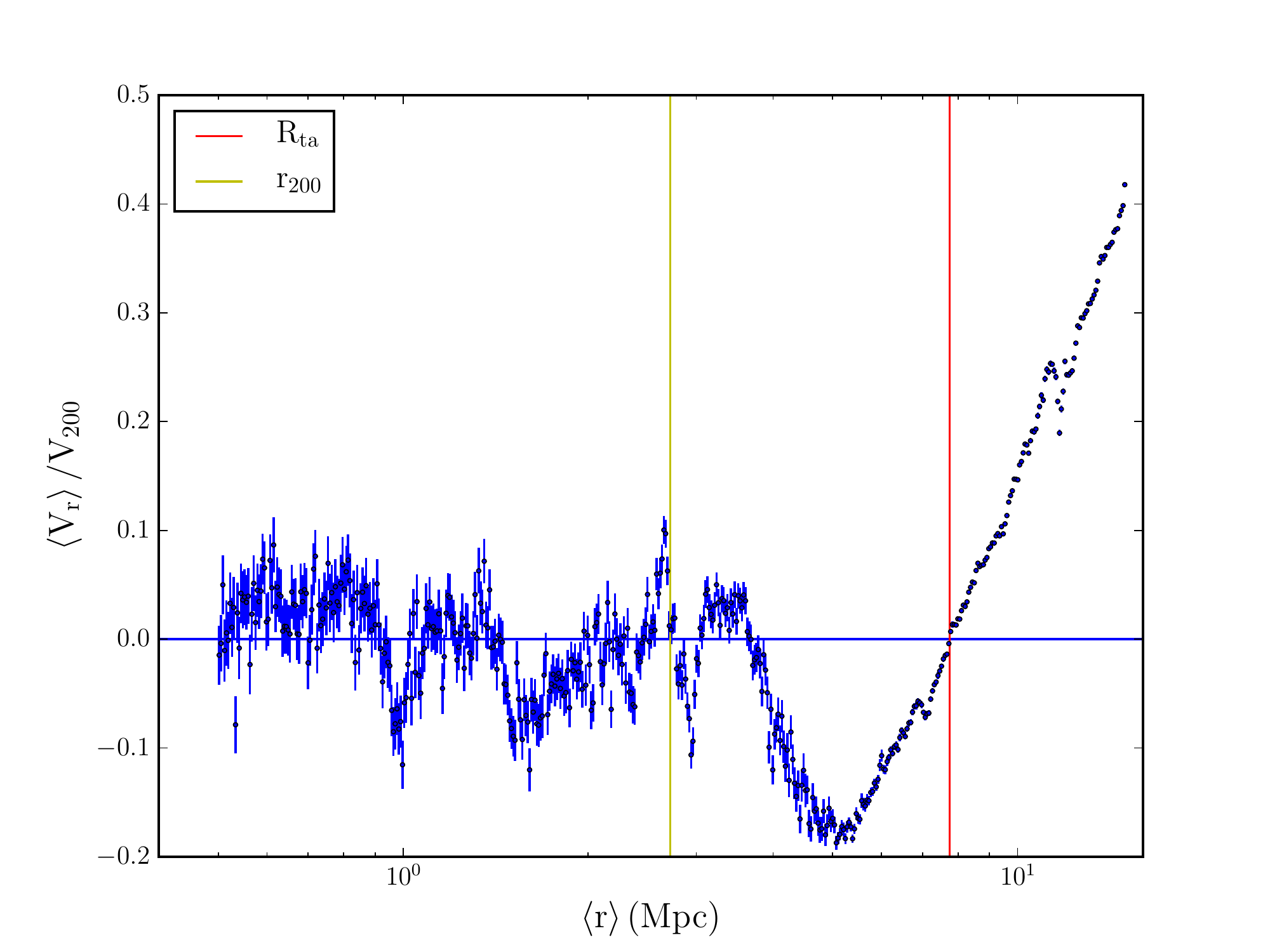}

\caption{Spherically-averaged radial velocity profiles of two dark matter halos, one from TDSS (left panel) with $ M_{200} = 2.15\times 10^{15} {\rm M_\odot}$, and one from Illustris-TNG (right panel) with $ M_{200} = 6.63\times 10^{14} {\rm M_\odot}$. Error bars indicate the $\langle V_r \rangle$ uncertainty (standard error of the mean) in each shell.  In both panels, the red vertical line indicates the turnaround radius.}
\label{Fig. 1.}
\end{figure*}

\begin{figure*}[htb!]
    \includegraphics[width=1.07\columnwidth]{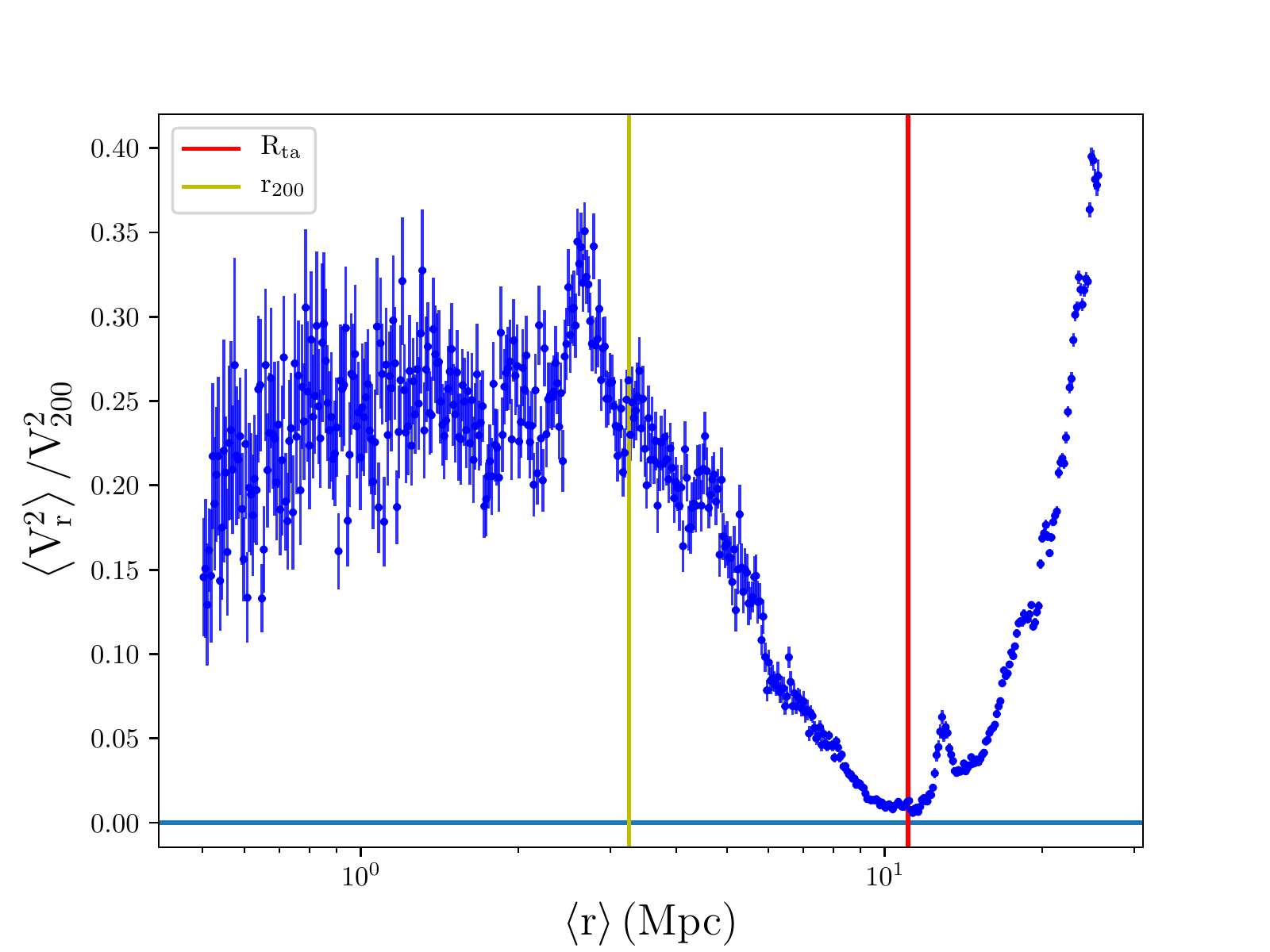}
    \includegraphics[width=1.07\columnwidth]{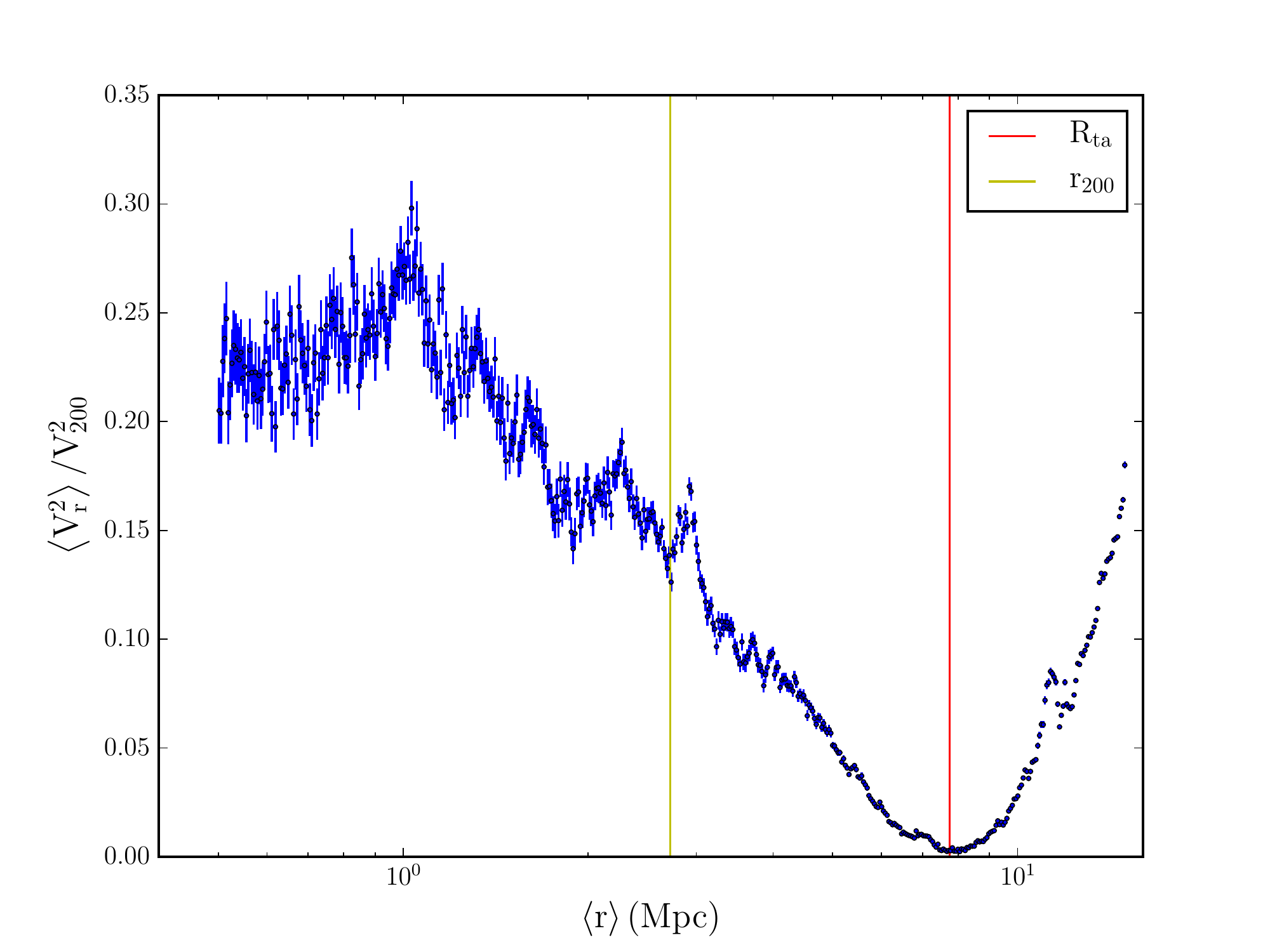}
\caption{Spherically-averaged profiles of the square of radial velocity, for the same objects as the corresponding panels of Fig.~\ref{Fig. 1.}. Error bars indicate the $\langle V_r^2 \rangle$ uncertainty (standard error of the mean) in each shell.  In both panels, the red vertical line indicates the turnaround radius, derived from the $\langle V_r \rangle$ profiles of Fig.~\ref{Fig. 1.}.}
\label{Fig. 2.}
\end{figure*}

   An important difference of the turnaround radius $R_{ta}$ from other halo boundary definitions, e.g. the virial radius or the splashback radius \citep{DK2014, splashback} is that $R_{ta}$ is kinematically identified\footnote{See however \cite{D2017} for an algorithm for the kinematic identification of the splashback radius. The ``static mass'' boundary of \citet{Cuesta} is also kinematically identified. See \S \ref{section discussion} for a detailed discussion of the differences between these different kinematically-relevant halo boundaries.}. For this, we use the radial profile of $\langle V_r \rangle$ (the average radial velocity around a halo center of particles residing within a spherical shell). In our analysis, spherical shells are logarithmically spaced, with the outer radius of each shell for Illustris-TNG being 0.68\% (0.78\% for TDSS) larger than the inner radius of the shell. The obtained value of $\langle V_r \rangle$ in each shell is then assigned to the average distance of all particles in that shell from the halo center. The uncertainty of $\langle V_r \rangle$ in each bin is taken to be the standard error of the mean.

   The radial profile of $\langle V_r \rangle$ exhibits a common pattern for every halo: beyond some distance from the halo center, matter follows the Hubble flow, making the average radial particle velocities positive thereafter. This typical behavior is shown in Fig.~\ref{Fig. 1.}, where radial velocity profiles are plotted for one halo from each simulation. Different halos, even in the same mass range, can feature different behavior in their spherically-averaged radial velocities in their inner and intermediate parts, depending on their relaxation state and merging history \citep{Cuesta}; however, all show a clear transition to positive $\langle V_r \rangle$ at their outskirts. The spatial location of this transition is the turnaround radius, $R_{ta}$ and in $\langle V_r \rangle - \langle r \rangle$ plots it is depicted by a vertical red line. 

   From each velocity profile, a value of $R_{ta}$ is obtained as follows: starting at very large radii and moving inwards,  the first crossing of ${\langle V_r \rangle=0}$ is located. The two points in the profile  straddling that first crossing are identified; the intersection of the line defined by these two points with the $r-$ axis is $R_{ta}$. 
   
   As discussed in the previous section, the value of $R_{ta}$ does exhibit some sensitivity to the selected centre of the structure. For this reason, once a value of $R_{ta}$ is estimated, the location of the center of the structure is reevaluated: the center of mass of all dark matter particles within a distance  $R_{ta}$ from the (original) structure center is calculated; then, the spherically-averaged radial velocity profile is recalculated with respect to that center of mass, and a new value of $R_{ta}$ is evaluated as above. This process is repeated until the value of $R_{ta}$ converges within $500$ kpc, or until five iterations are performed. In the latter case, the structure is flagged.
 
Once we have estimated the locations of the centers of mass and $R_{ta}$, we address the substructure issue. For each structure, all neighboring structures with centers within $R_{ta}$ are identified. Of these (selected neighbors and original structure) the object with the largest $M_{200}$ is labeled as "structure" and retained. All the others are labeled as "substructure" and removed from further consideration.  Our substructure cleaning algorithm did not find any substructure at turnaround included in the TDSS halos, and it identified and removed six substructrues from the Illustris-TNG halo sample.

\section{Results} \label{section 4}
 \subsection{Can a single turnaround radius meaningfully characterize a realistic 3D structure?}

In an expanding universe that is on large scales homogeneous and isotropic, the turnaround radius is always a well-defined scale, even for a structure that exhibits strong deviations from spherical symmetry in both its mass distribution and its kinematics. The collapsed center of the structure will not be expanding; eventually, large enough scales will join the Hubble flow and will expand. Thus, the spherically-averaged radial velocity profile {\em will} pass through zero, and the distance from the center where this occurs can always be defined as the "turnaround scale". Whether this scale meaningfully characterizes a strongly non-spherical collapsed structure is however a different, non-trivial question. One could imagine that for a strongly non-spherical structure, particles in different directions turn around at different distances from the center. Individual particles within the ``turnaround'' shell could thus exhibit significant inwards or outwards motions in different directions that cancel out on average, a behavior similar to what is usually encountered at the centers of structures.

\begin{figure}[htb!]
    \includegraphics[width=1.07\columnwidth]{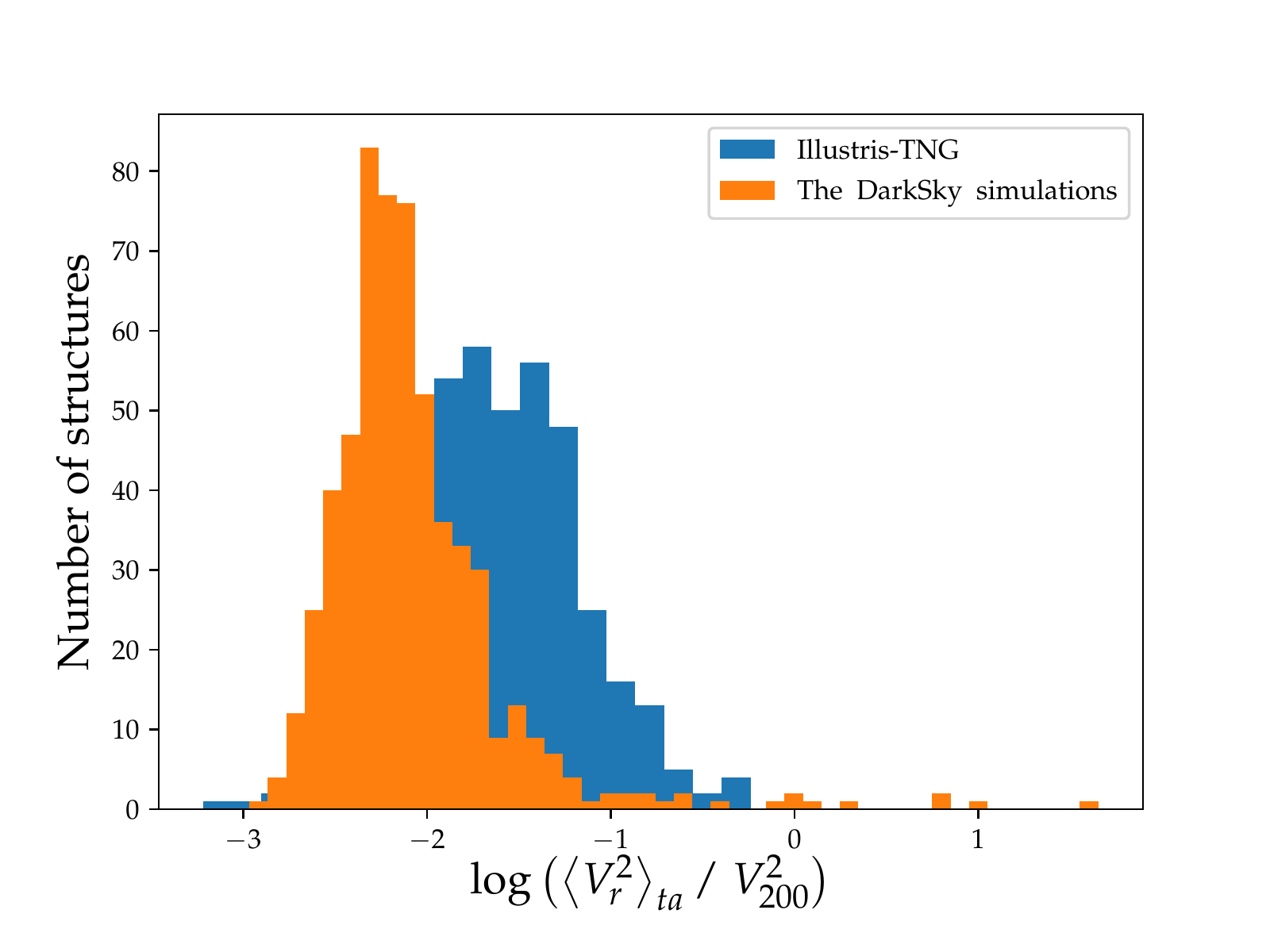}
    \includegraphics[width=1.07\columnwidth]{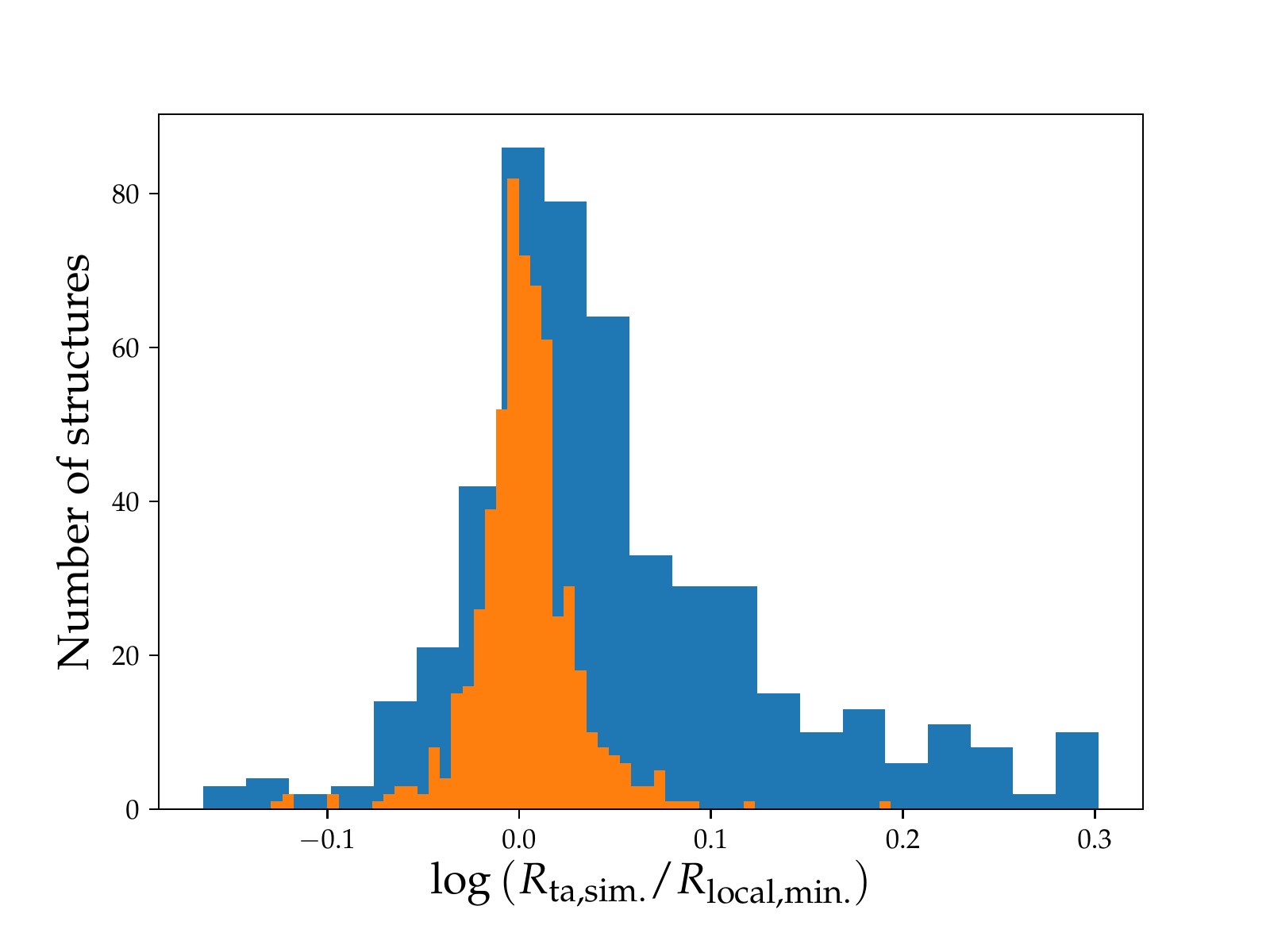}
 \caption{Upper panel: distribution of $\langle V_r^2\rangle$ in the turnaround shell, in units of $V^2_{200}$,  for halos in TDSS (orange histogram) and Illustris-TNG (blue histogram): the turnaround shell is indeed characterized by very small individual radial velocities of all its particles. Lower panel: distribution of the logarithm of the ratio of the turnaround radius of a structure over the radius of the shell where $\langle V_r^2\rangle$ is minimum in that same structure. }
 \label{velocity distributions}
 \end{figure}

\begin{figure*}
\begin{multicols}{3}
    \includegraphics[width=6.3cm]{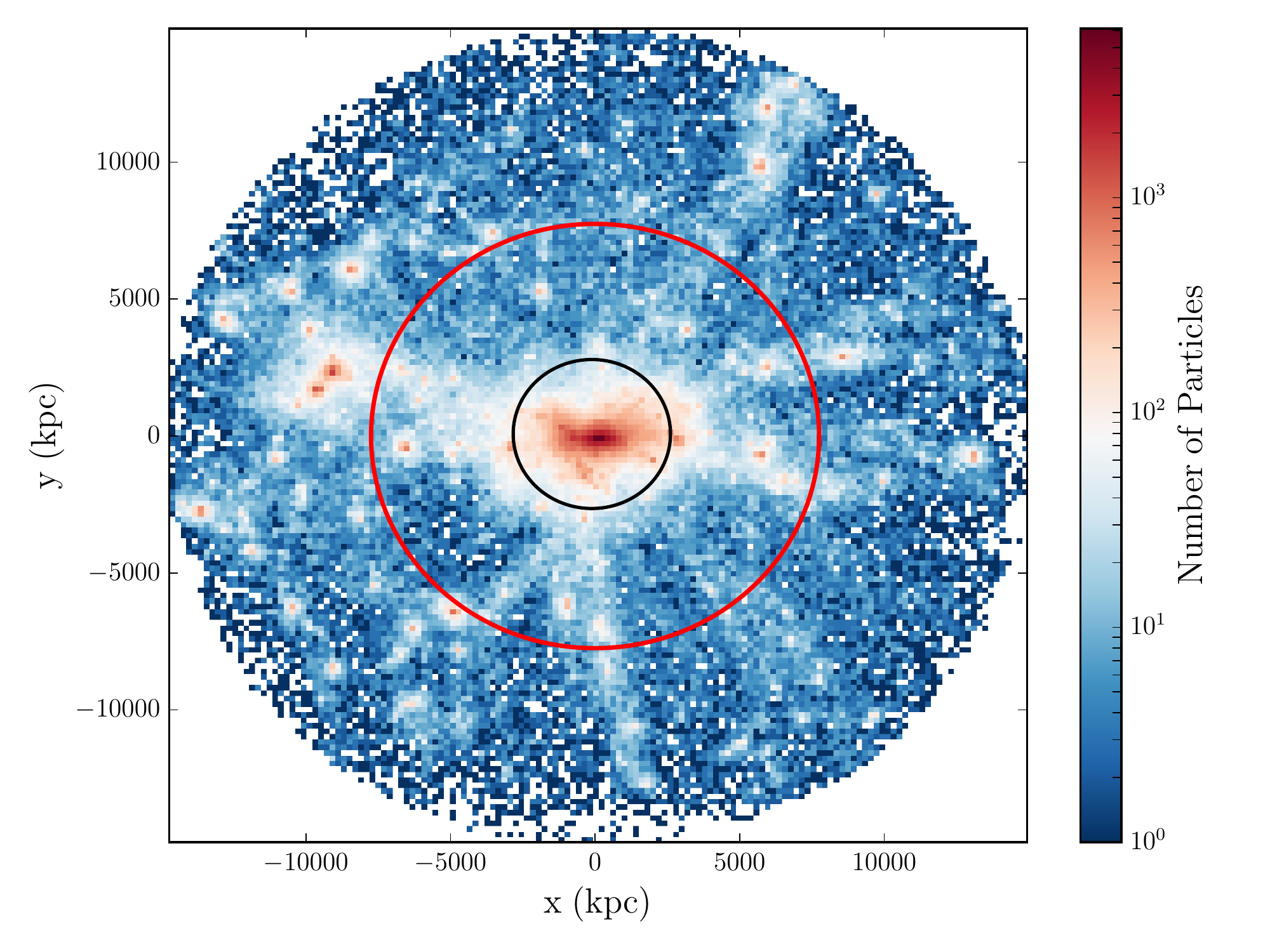}\par 
    \includegraphics[width=6.3cm]{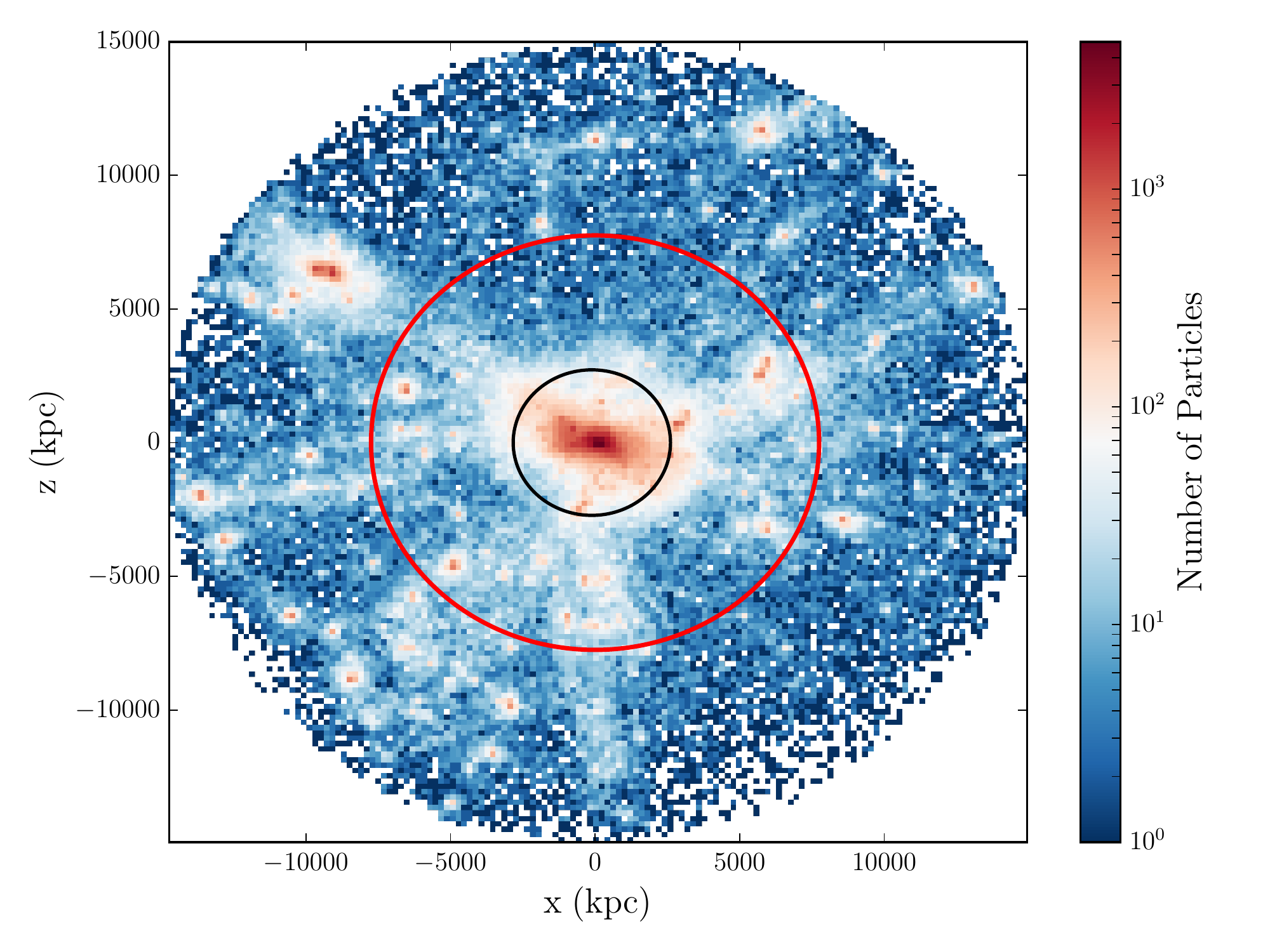}\par
    \includegraphics[width=6.3cm]{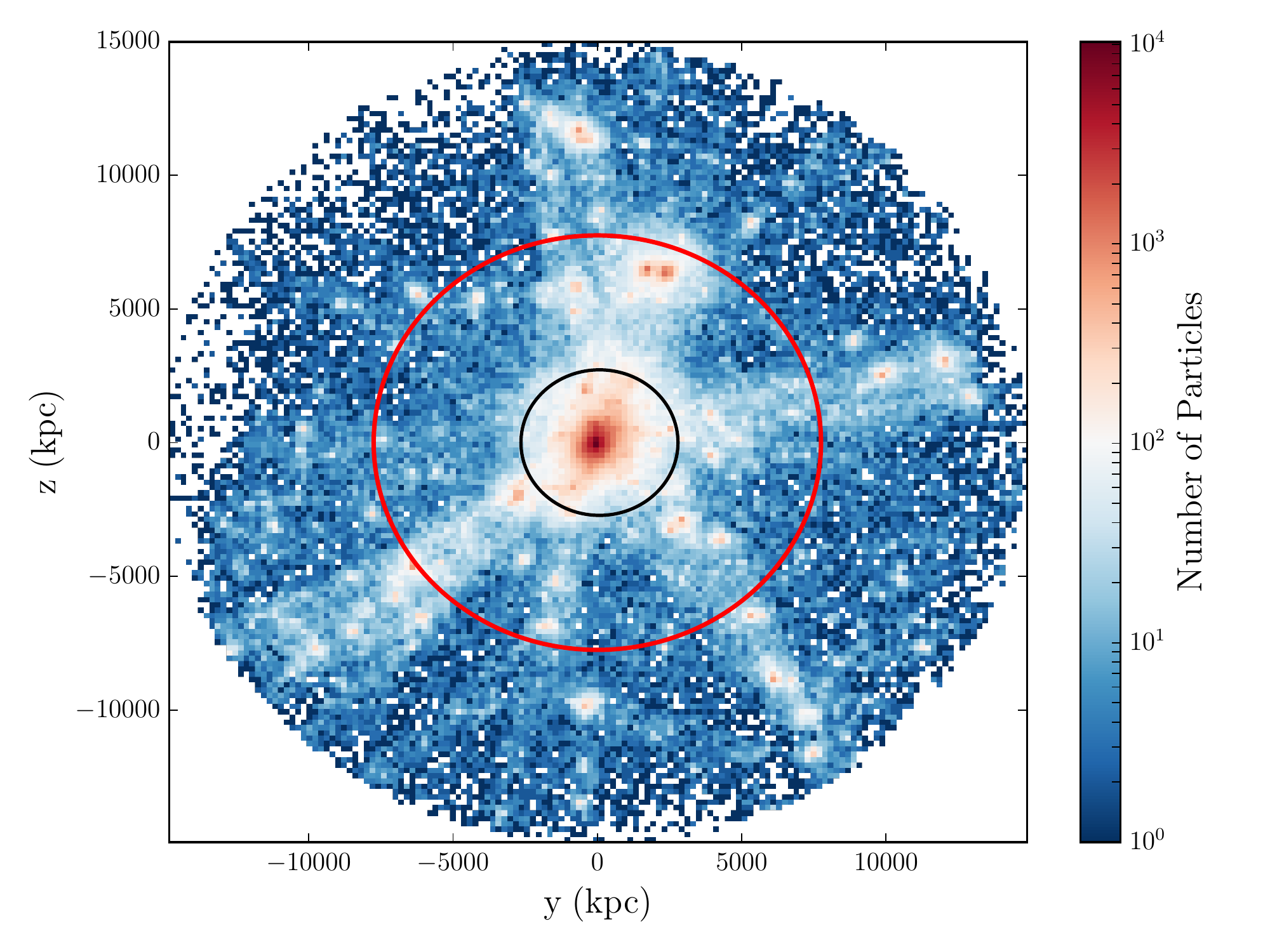}\par

\vspace{0.1cm}
    
    \includegraphics[width=6.3cm]{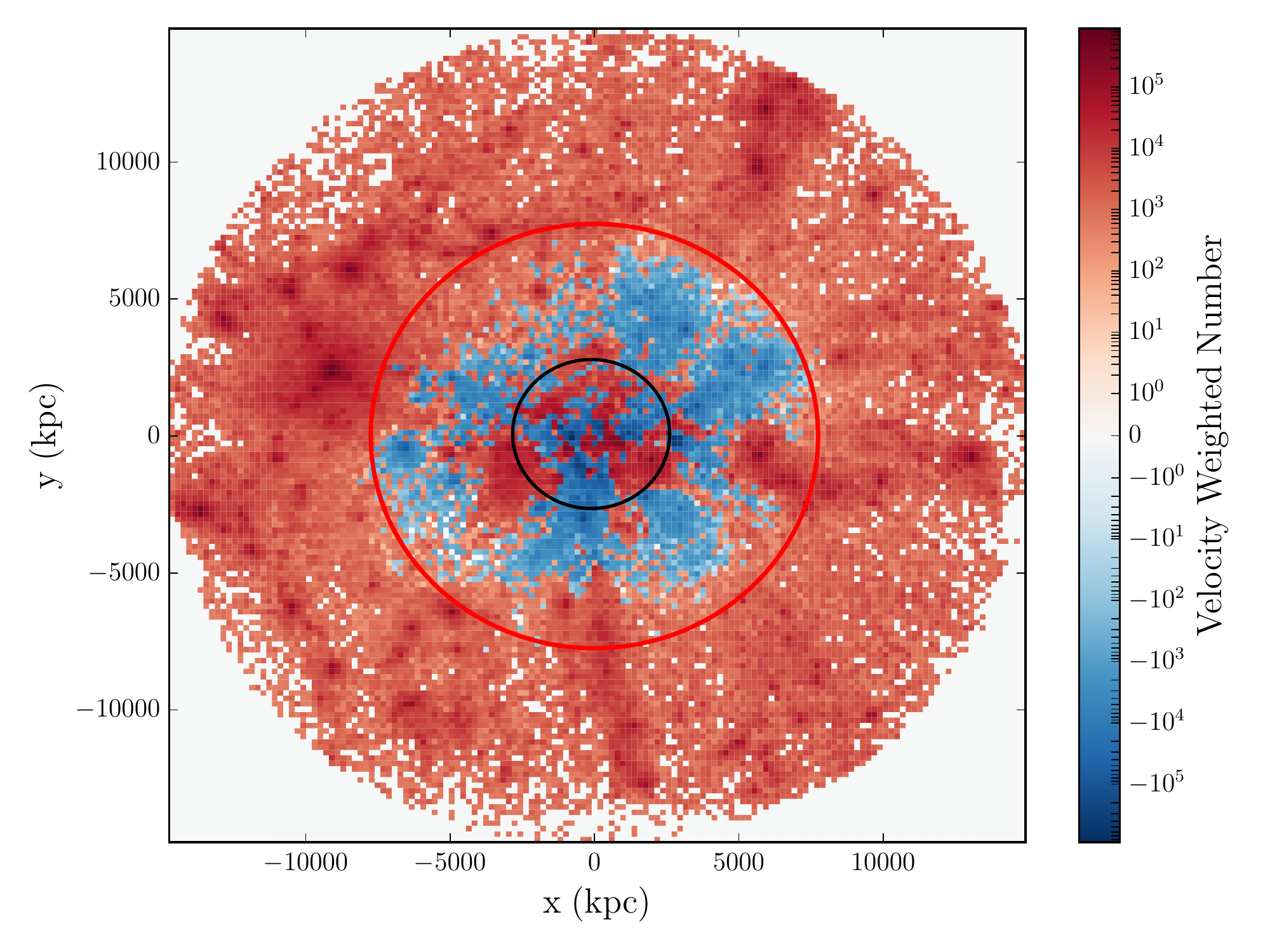}\par 
    \includegraphics[width=6.3cm]{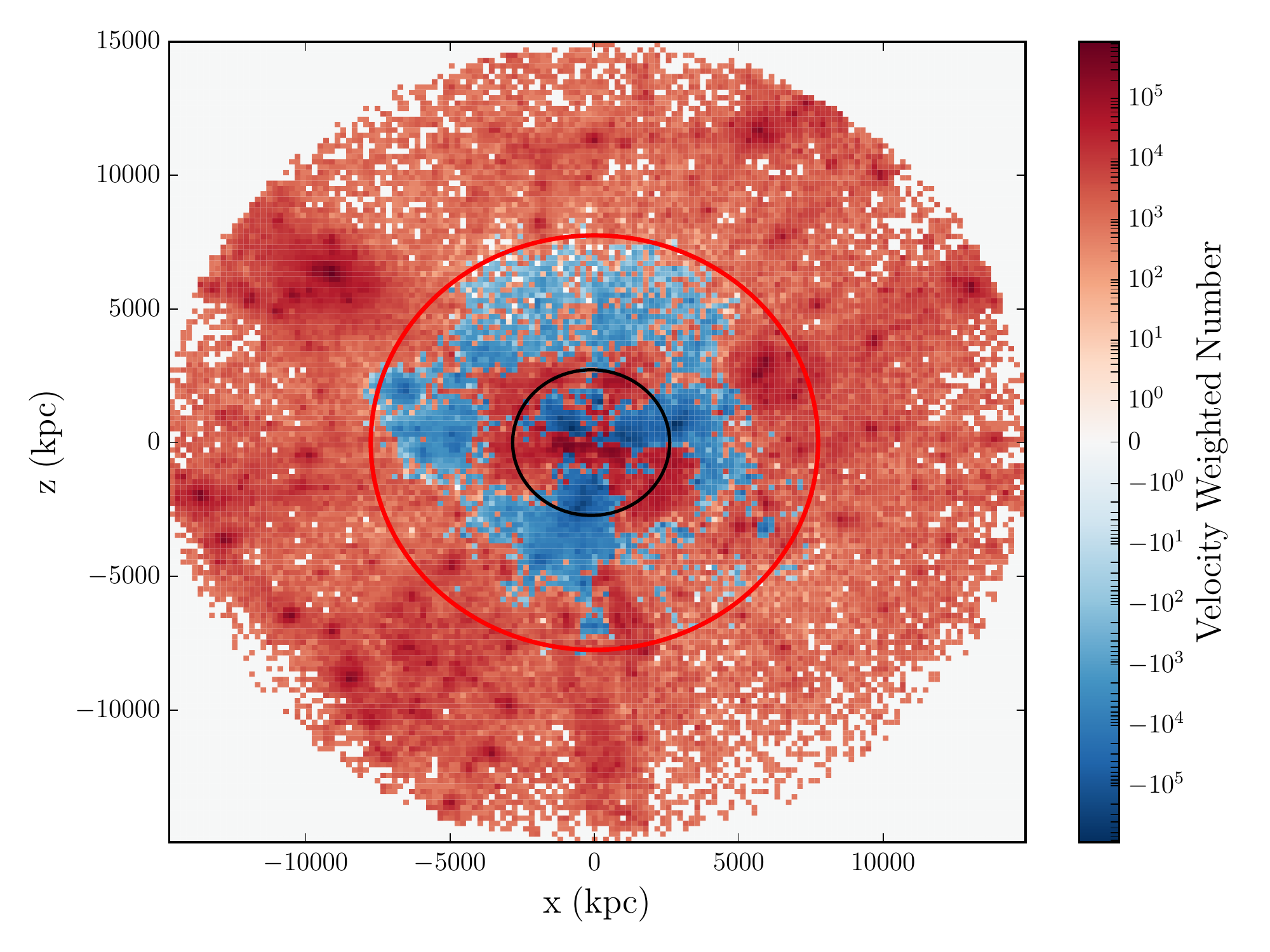}\par
    \includegraphics[width=6.3cm]{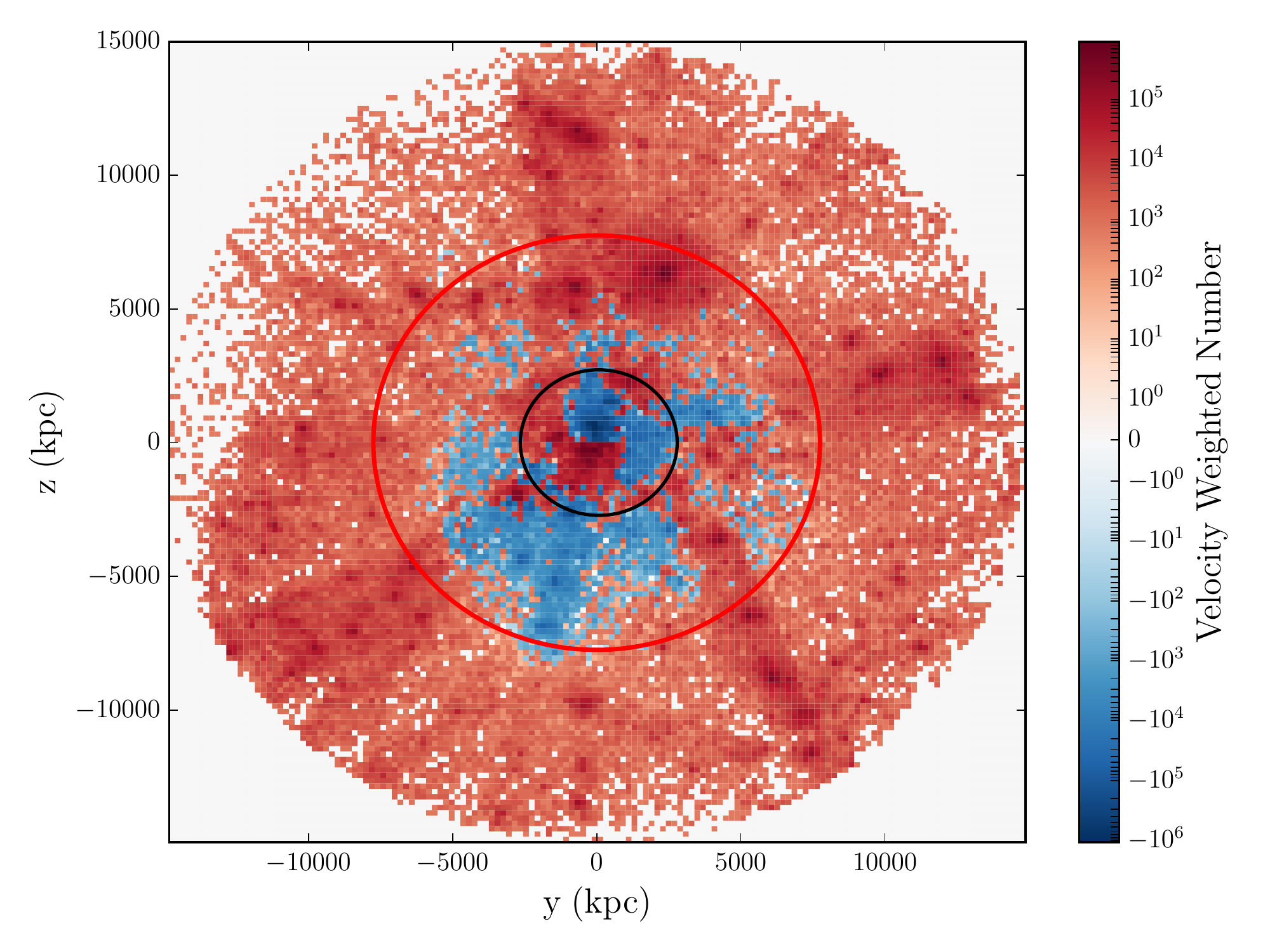}\par

\vspace{0.1cm}

    \includegraphics[width=6.3cm]{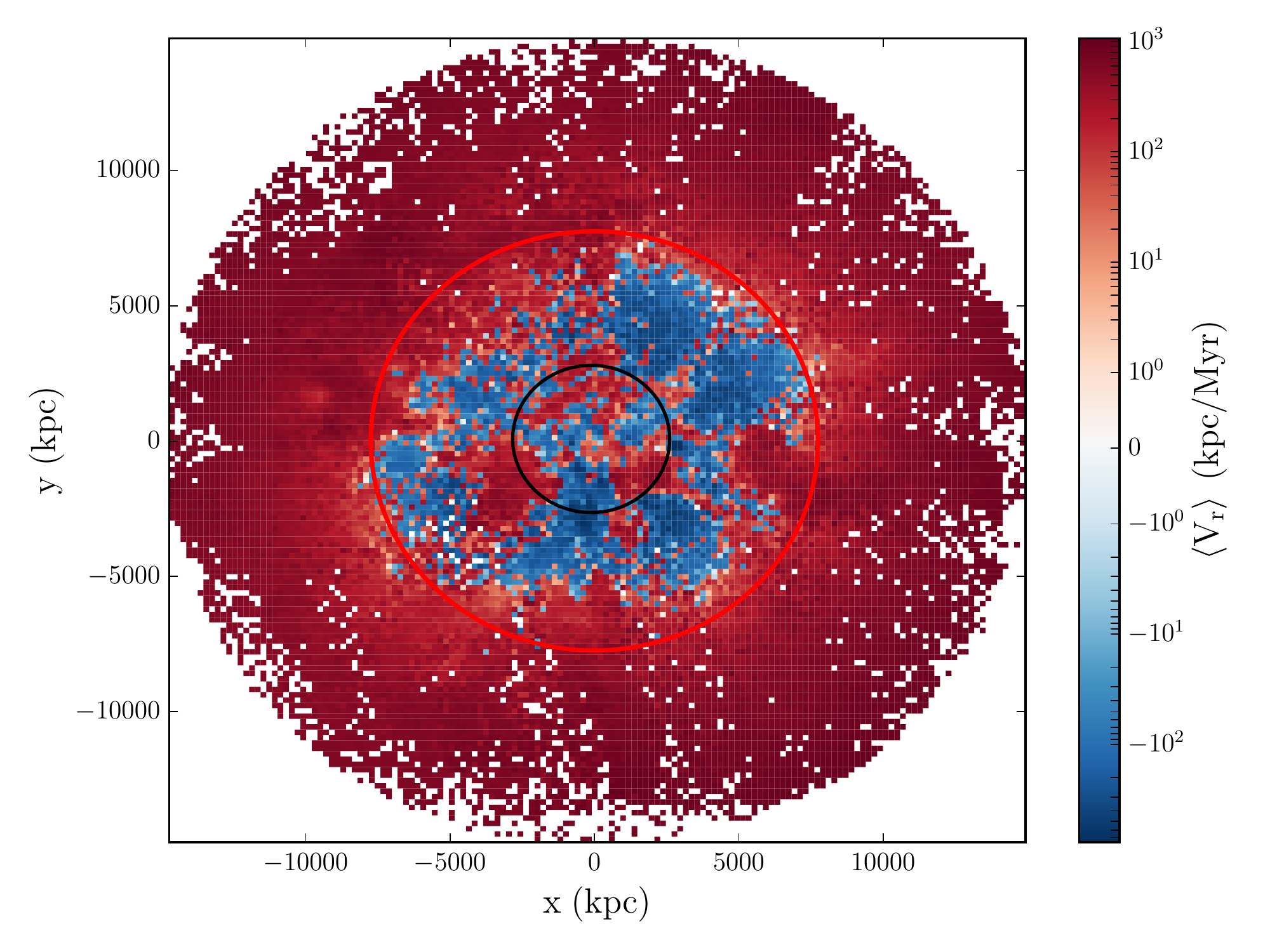}\par     \includegraphics[width=6.3cm]{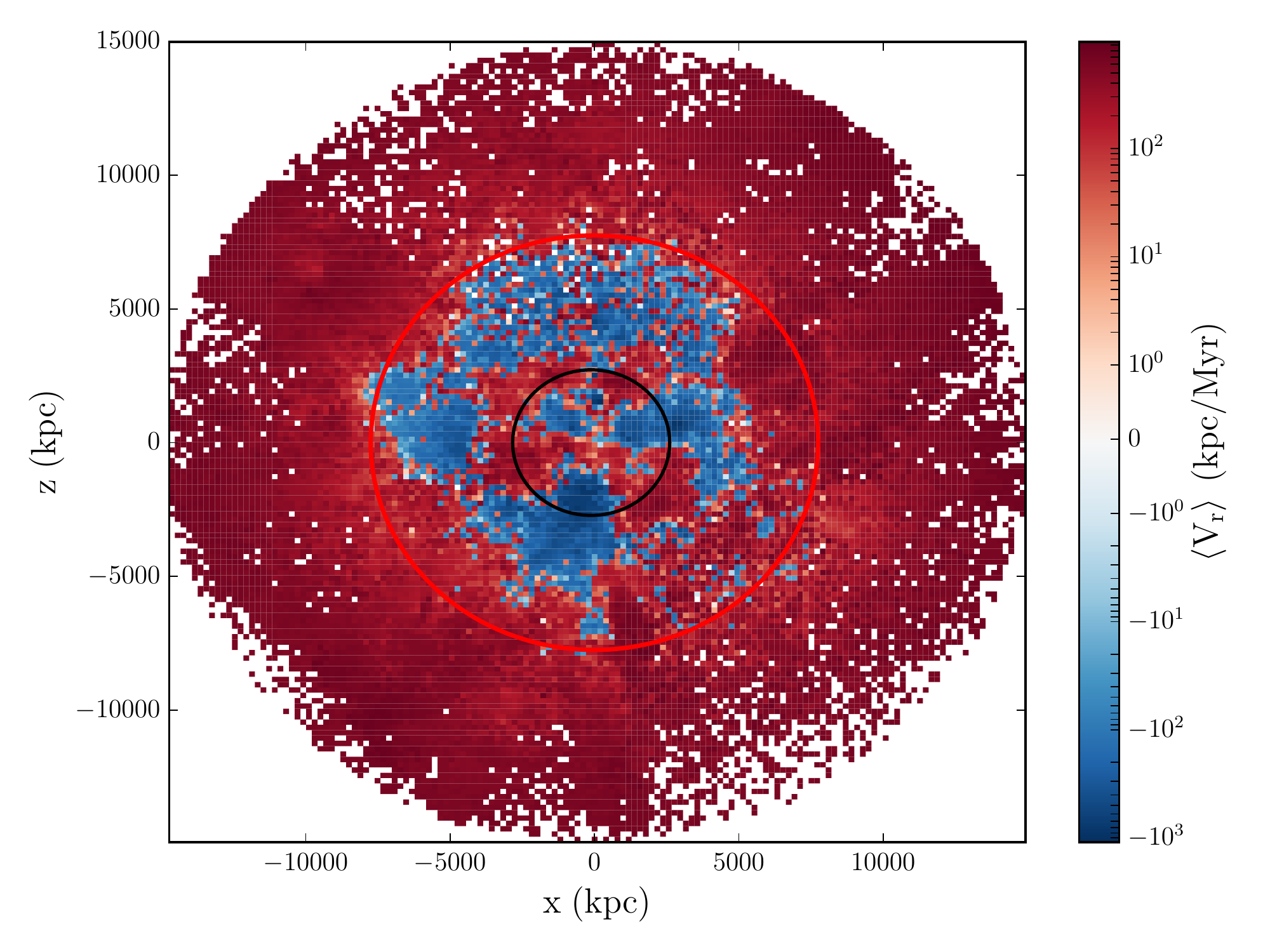}\par
    \includegraphics[width=6.3cm]{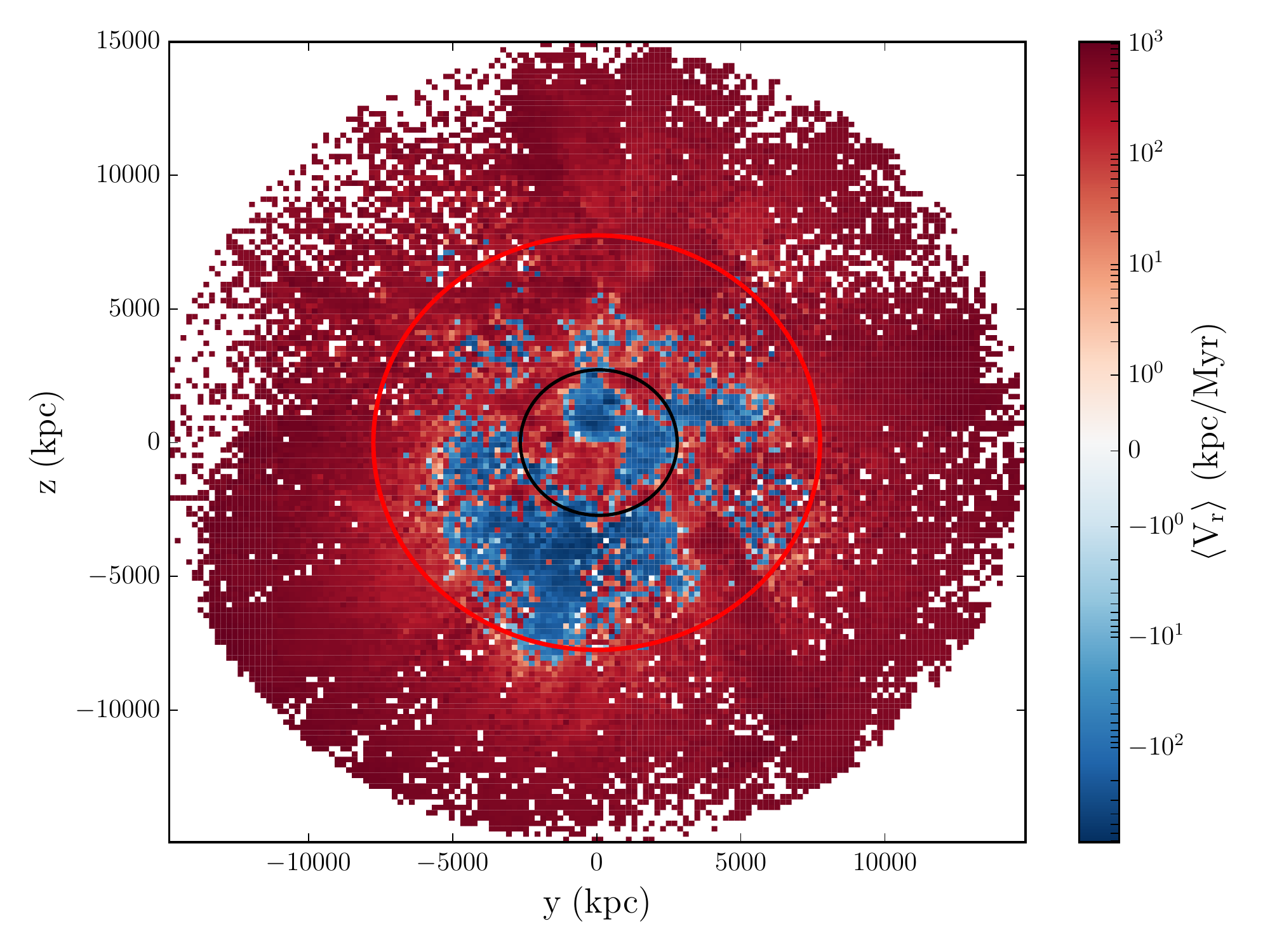}\par
    \end{multicols}
\caption{2D projections of the ${\rm M_{200}} = 6.63 \times 10^{14}\, {\rm M_\odot}$ structure from Illustris-TNG depicted in the right-hand panels of Figs.~\ref{Fig. 1.} and \ref{Fig. 2.}. Left column: number of particles (representing column density). Middle column: sum of particle radial velocities (representing radial-velocity--weighted column density). Right column: average radial velocity (representing projections of the radial velocity field).  Black circles indicate the location of $R_{200}$ and  red circles the location of $R_{ta}$. In the midle and right columns, red hues correspond to outflow and blue hues to infall.}

\label{Figure Weighted projections}
\end{figure*}
In order to test for such a scenario, we have constructed the spherically-averaged profiles of the {\em square} of the radial velocity. In a spherically symmetric structure, the turnaround radius is where all particles stand still, and the square of the radial velocity will become zero at the turnaround scale. If a single turnaround scale indeed characterizes realistic structures in our simulations, it should be similarly imprinted in the spherically-averaged profiles of $V_r^2$ as a pronounced global minimum, very close to zero compared to typical squared velocities in the halo. This is indeed the case as we can see in Fig.~\ref{Fig. 2.}.  The two profiles correspond to the same structures as the $\langle V_r \rangle$ profiles of Fig.~\ref{Fig. 1.}. 
This behavior of low/minimum $\langle V_r^2\rangle$ in the turnaround shell is typical for halos in the two simulations. We demonstrate this in Fig.~\ref{velocity distributions}. 
The upper panel shows the distribution of $\langle V_r^2\rangle$ at turnaround normalized to $V_{200}^2$ of each halo, and demonstrates that the radial velocity dispersion in the turnaround shell is {\em small} for the vast majority of halos in our sample. The lower panel shows the distribution of the ratio between the turnaround radius, and the radius where $\langle V_r^2\rangle$ is minimum, demonstrating that the two are very close in all halos.  We conclude that the turnaround radius obtained from the spherically-averaged radial velocity profile is indeed a kinematically and dynamically meaningful scale that can describe the boundary of halos in N-body simulations, despite the very strong deviations of the mass distribution of these halos from spherical symmetry. 

In order to visualize the origin of this result, we plot in Fig.~\ref{Figure Weighted projections} two-dimensional projections of the ${\rm M_{200}} = 6.63 \times 10^{14}\, {\rm M_\odot}$ structure from Illustris-TNG depicted in the right-hand panels of Figs.~\ref{Fig. 1.} and \ref{Fig. 2.}. The turnaround radius is shown with a red circle, while a black circle indicates the location of $R_{200}$.  The left column visualizes the halo column density along the three different coordinate axes, by plotting the number of dark matter particles that are projected in each bin on the plane. The mass distribution of the halo is very anisotropic on the turnaround scale. The turnaround scale itself is not clearly identifiable in the column density projections. Despite the presence of substructure, there is a very pronounced mass concentration at the center of the halo that represents a significant fraction of the total mass enclosed by the turnaround radius. 

The middle and right columns on Fig.~\ref{Figure Weighted projections} represent different renderings of radial velocities. The middle column visualizes the radial-velocity--weighted dark matter column density, by plotting the sum of particle radial velocities in each bin on the plane. Red hues correspond to outflow and blue hues to infall. The radial velocity field itself is shown in the right column, where the color in each bin on the plane corresponds to the average radial velocity of particles projected in that bin. Here the turnaround scale is immediately identifiable as a kinematic boundary of the structure, even without the guidance of the red circle. Interestingly, at the turnaround scale the collapse/expansion is much less anisotropic than the mass distribution, and much less anisotropic than the veolicty structure near the center. At the center, the collapse/bounceback of the dark matter is homogeneous only to the extent that the in-falling satellite halos have been integrated to the central halo. In contrast, at the turnaround scale the dynamics appear to be dominated by the central mass concentration. The gravitational potential at these large distances is then much closer to a central potential, and this is reflected in the considerably reduced anisotropy in the collapse/expansion around the kinematic boundary of the object. 

 \subsection{How does the size of the turnaround radius of 3D structures compare to the predictions of the spherical collapse model for objects of the same enclosed mass?}
 
\begin{figure}[htb!]
    \centering
    \includegraphics[width=1.07\columnwidth]{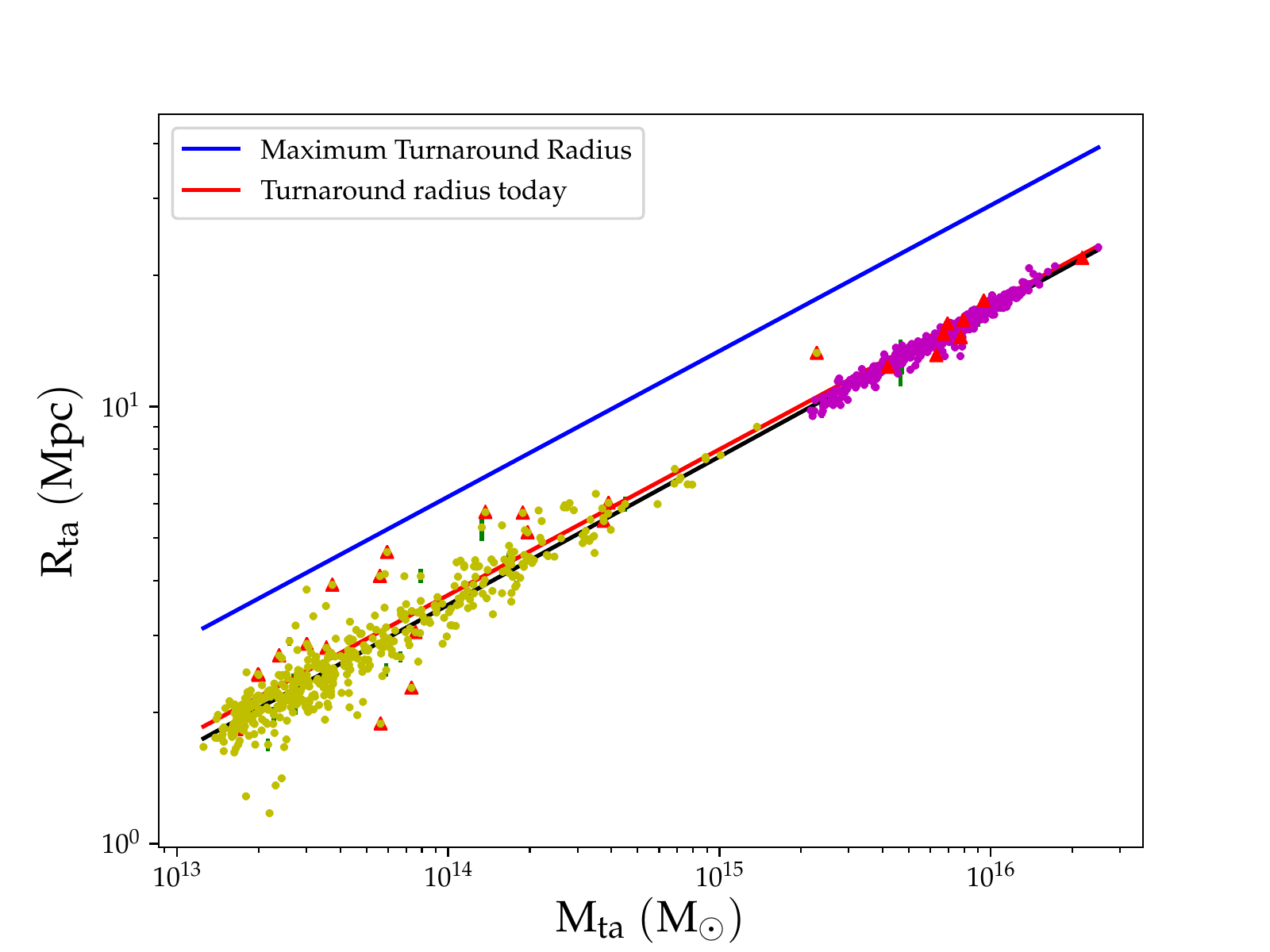}
\hfill
\caption{Turnaround radius as a function of enclosed mass. Yellow points:  Illustris-TNG halos; magenta points:  TDSS halos. The two simulations use different cosmological parameters;  TDSS data points have been renormalized to scale with cosmological parameters of Illustris-TNG. Blue/red lines:theoretical predictions of the spherical collapse model for maximum (asymptotic late-time)/present-time values  of turnaround radius. Black line: power-law fit (slope: $0.338 \pm 0.001$).}
\label{Fig. 5.}
\end{figure}

 We address this question in Fig.~\ref{Fig. 5.}, where we plot the turnaround radii for the halos we have analyzed as a function of the mass enclosed within that radius. All halos were cleaned from substructure with the method described in \S \ref{section 3}.
Because the two simulations do not have exactly the same cosmological parameters, we have rescaled all $R_{ta}$ in TDSS by $(\rho_{\rm Illustris-TNG}/\rho_{\rm TDSS})^{-1/3}$ where $\rho_{\rm Illustris-TNG}$ and $\rho_{\rm TDSS}$ are the turnaround densities predicted by spherical collapse for $z=0$ for the Illustris-TNG and TDSS cosmological parameters, respectively. 
 There is an extremely tight correlation between turnaround radius and enclosed mass. A power-law fit (black line) yields a scaling very close to $R_{ta} \sim M_{ta}^{1/3}$ (best-fit slope $0.338 \pm 0.001$), as expected from structures of comparable average density.

  In the same figure we overplot with the blue line the scaling with mass of the maximum (asymptotic late-time) turnaround radius \citep{Pavlidou_Tomaras}, the normalization of which only depends on $\Lambda$. The red line represents the scaling predicted by the spherical collapse model for the turnaround radius of structures of all masses at the present cosmic time \citep{Tanoglidis2015}, and it is in striking agreement with the turnaround radii of simulated halos, despite the strong deviations of the latter from spherical symmetry. Halos indicated with triangles in this plot correspond to cases where the location of the center of mass of the halo failed to converge within 500 kpc after five iterations. Such cases of ambiguous halo center account for a significant fraction of the outliers from the average scaling and the predictions of the spherical collapse model. 
 
Despite the scatter that is pronounced for halos of lower virial masses, the bulk of the structures lie well below the upper bound indicated  by the blue line. However, there are several structures that are close to it, and it is entirely conceivable that additional inaccuracies in the observational determination of the turnaround radius and the enclosed mass of structures may result in individual measurements that are in violation of the bound.  Our results are thus consistent with the findings of \citet{LY16} that any single observation in violation of the bound is not directly at odds with $\Lambda$CDM, given observational uncertainties and realistic structure-to-structure scatter. However, unless a systematic bias is present in observations, the {\em average} behavior of the structures should track well the predictions of spherical collapse; if anything, simulated halos exhibit a slight bias towards {\em smaller} values of the turnaround radius compared to the expectations from spherical collapse for structures of the same enclosed mass (i.e. a bias towards a higher value of the average "turnaround density", see also \S \ref{density-distribution}). 

 \subsection{Is the average turnaround density of realistic 3D halos independent of halo mass at a given cosmic epoch? }\label{density-distribution}
 
 Given that $R_{ta}$ was shown in Fig.~\ref{Fig. 5.} to scale with enclosed mass as $M_{ta}^{1/3}$, it is clear that a characteristic average density for structures defined on turnaround scales (a "turnaround density" $\rho_{ta}$) does exist. The good agreement of the normalization of the $R_{ta}-M_{ta}$ scaling with the predictions of the spherical collapse model imply that this characteristic density is indeed close to the predicted density of a single spherical structure in an otherwise unperturbed Universe turning around today, $\rho_{ta,sph.coll.,z=0}$. 
 
 In this section, we investigate the behavior of the distribution of $\rho_{ta}$ in simulated halos with $M_{200}>10^{13}M_\odot$. This is important for two reasons. First, in the spherical collapse model, it is the evolution of $\rho_{ta}$ with redshift that probes cosmological parameters on turnaround scales (\citealp{Tetal16}; Pavlidou et al. 2019, in prep.). Second, if indeed the distribution of $\rho_{ta}$ is strongly peaked around the characteristic value, this could provide a straight-forward way to estimate the turnaround radius of a structure based on its density profile alone, just as the "virial radius", $R_{200}$, is identified in simulations and observations as the radius of a given density contrast with the  mean matter-density of the Universe.  We stress however that, unlike $R_{200}$ which is obtained {\em demanding} that the enclosed structure has an average density of 200$\times$ the average matter-density of the Universe at the same cosmic epoch, $R_{ta}$ is derived from each structure's velocity profile, with no {\it a priori} constraint on the enclosed matter density. That such a characteristic density does appear to exist  is a physical result of the dynamics of the problem rather than of our halo-finding algorithm.

\begin{figure}[htb!]
    \centering
    \includegraphics[width=1.07\columnwidth]{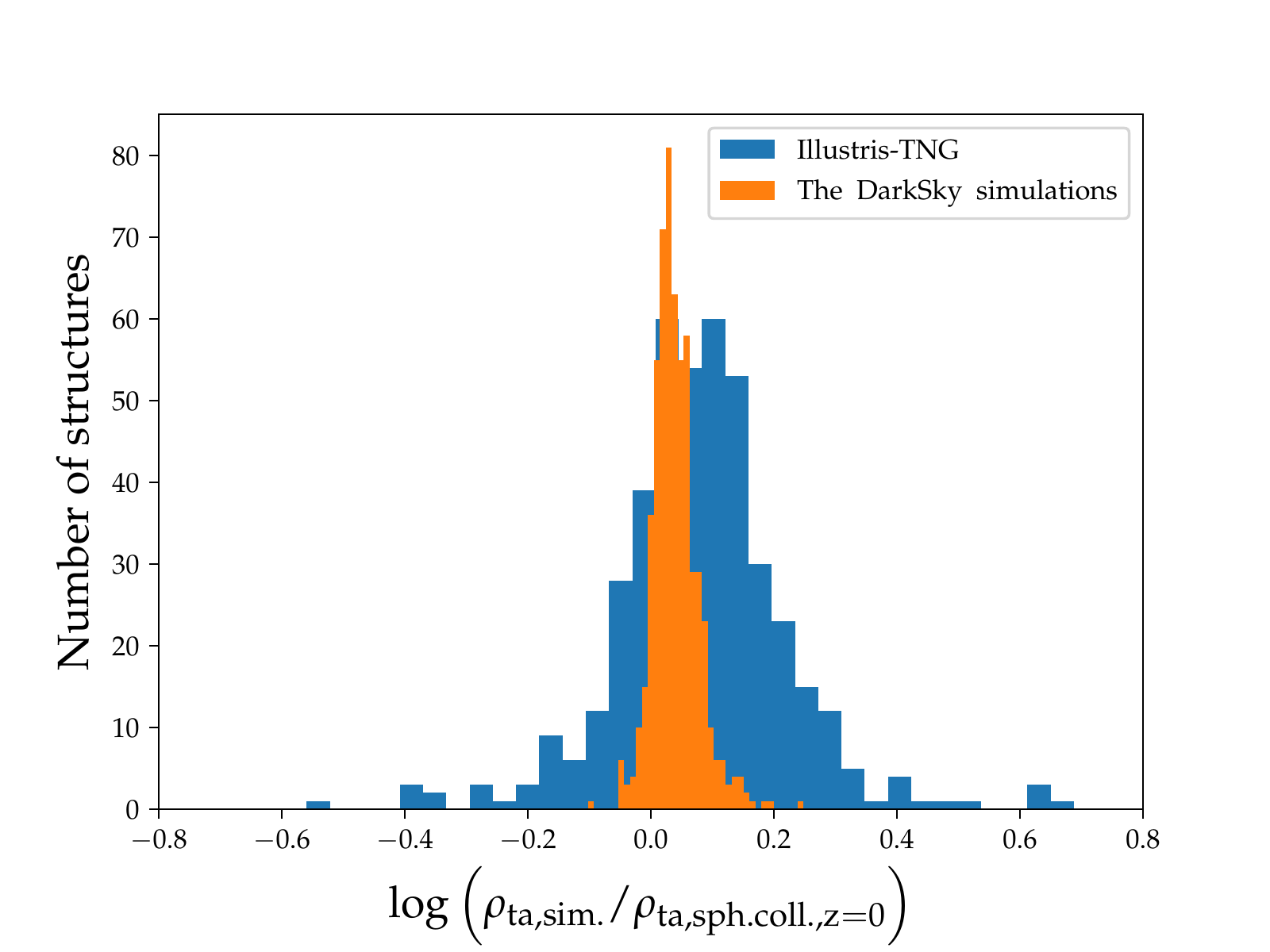}
    \caption{Distribution of the logarithmic difference between average matter density within $R_{ta}$ and the characteristic value predicted by the spherical collapse model {\em for all structures at z=0} independently of their mass. The orange histogram corresponds to (higher mass) TDSS halos, while the blue histogram to (lower mass) Illustris-TNG halos.}
\label{Fig. densities}
\end{figure}

 To this end, in Fig.~\ref{Fig. densities} we plot the distribution of the logarithm of the ratio between $\rho_{\rm  ta, sim}$ measured in simulations, and the value $\rho_{\rm ta,sph.coll.,z=0}$ prediced by spherical collapse.
 A value of $\log(\rho_{\rm ta, sim}/\rho_{\rm ta,sph.coll.,z=0}) = 0$ corresponds to perfect agreement with spherical collapse predictions. Different colors correspond to different simulations and, due to the difference in the size the simulation boxes, different halo masses (as can be also seen in Fig.~\ref{Fig. 5.}). Both distributions are strongly peaked close to 0: TDSS halos have a median $\log \left( \rho_{\rm ta, sim}/\rho_{\rm ta,sph.coll.,z=0}\right)$ of 0.04, and a standard deviation of 0.04; Illustris-TNG halos have a median of 0.08 and a standard deviation of 0.15. The difference between halos in the two simulations are because of the difference in the masses of the halos we have analyzed in each simulation. The larger TDSS halos are in better agreement with spherical collapse predictions. For both simulations, $\rho_{\rm ta}$ in simulations is consistent with spherical collapse predictions within one standard deviation. There is however a systematic bias towards higher $\rho_{\rm ta}$  (lower $R_{ta}$ for a given enclosed mass) in simulations;  the bias decreases with increasing mass. The same trend can be seen in Fig.~\ref{Fig. 5.}.  

 \subsection{Is there a dependence of the turnaround radius on the shape of structures?}
 
Although the correlation between $R_{ta}$ and $M_{ta}$ is clearly significant, has the correct slope, and a normalization very close to the prediction of the spherical collapse model, it does feature appreciable scatter, especially for structures of lower masses, and a slight normalization offset.

We have already identified the ambiguity in the definition of the center of a turnaround structure as one of the sources of the scatter. Intuitively, one would argue that deviations of the matter distribution in a halo from spherical symmetry would also be a prime suspect as a cause for deviations of the turnaround radius from the predictions of the spherical collapse model. To quantify and evaluate this hypothesis, we construct a measure $\alpha$ of the asphericity of a halo on  turnaround scales. To this end, we calculate the principal moments of inertia $I_k$  with  $k=1,2,3$, and define $\alpha$ as 
\begin{equation}
  \alpha = \frac{I_{k,min}}{I_{k,max}}
\end{equation}
   so that a value of 1 corresponds to a sphere whereas a value 0 to a prolate/oblate object of infinitesimal thickness. 
  
  \begin{figure}[!t]
\includegraphics[width=1.05\columnwidth]{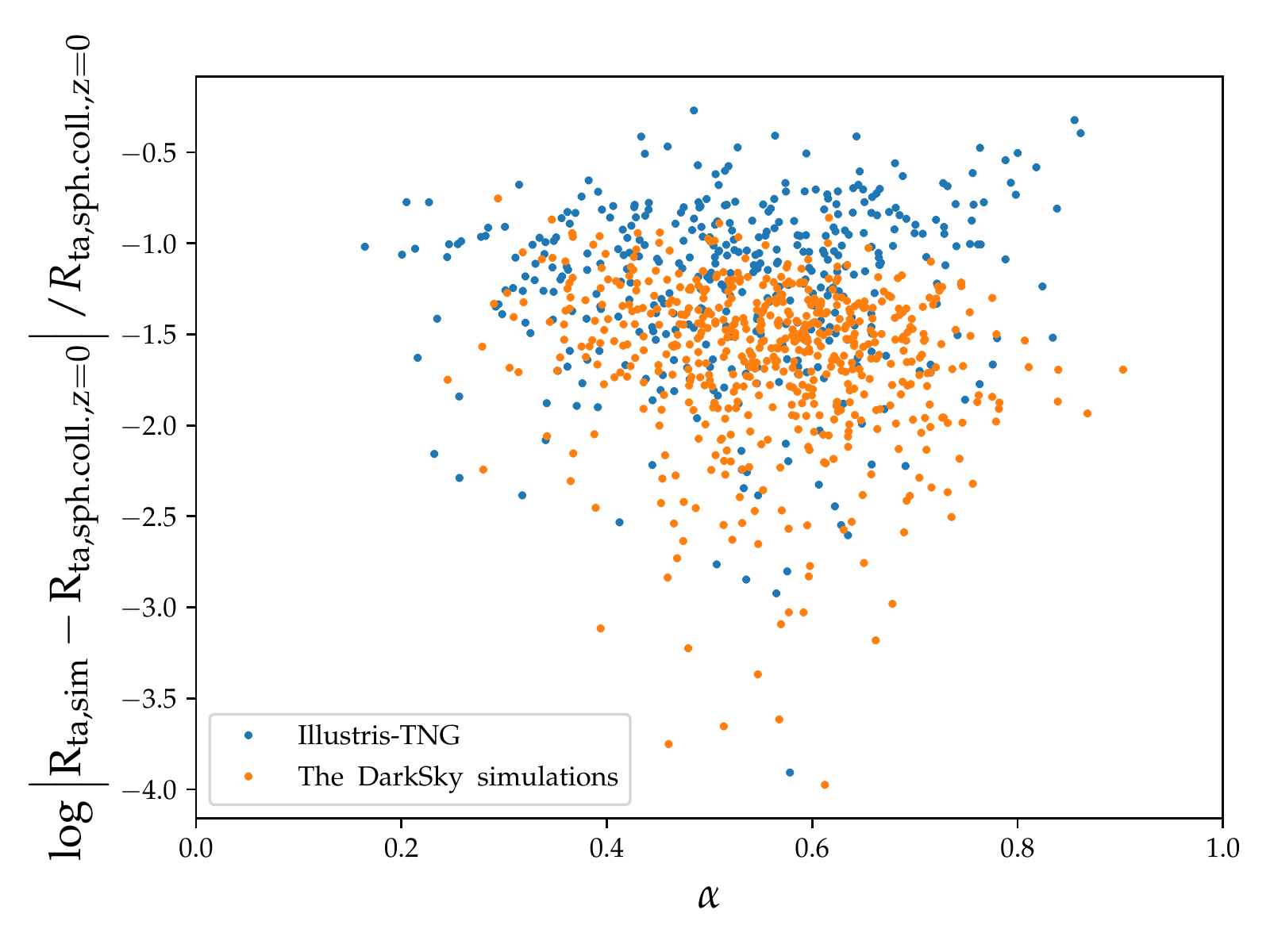}
\caption{Fractional deviation of $\rm{R_{ta}}$ from the value predicted by spherical collapse for a structure of identical enclosed mass at $z=0$, plotted against the asphericity parameter $\alpha$. Higher $\alpha$ correspond to halos closer to spherical symmetry. No correlation is seen for the smaller  Illustris-TNG halos (Spearman correlation coefficient: $0.01$; p-value: $4\%$) and a very weak although significant correlation is seen in the higher-mass TDSS halos (Spearman correlation coefficient: $ -0.16$; p-value: $8\times 10^{-6}$).}
\label{Fig. 6.}
\end{figure}

In Fig.~\ref{Fig. 6.} we plot the absolute fractional deviation of $R_{ta}$ from the corresponding value predicted by spherical collapse for a structure of the same mass against the asphericity  parameter $\alpha$. For the lower-mass Illustris-TNG halos (blue dots) no discernible trend is seen, despite the presence of a large range of shapes and appreciable deviations from spherical collapse: the dominant cause of these deviations in smaller halos is {\em not} their shape. This is confirmed by a formal correlation analysis, that yields a Spearman correlation coefficient of $0.01$ and a p-value of $4\%$: the correlation is extremely weak, in the opposite direction than intuitively expected, and not statistically significant.  For the larger TDSS halos the deviations from spherical collapse predictions are much smaller to begin with. Here some trend of decreasing deviations with decreasing asphericity (increasing $\alpha$) can be seen, and a correlation analysis confirms  this (p-value $8\times 10^{-6}$). However the trend is very weak (Spearman correlation coefficient $-0.16$). 

These results do not confirm the prediction of \citet{GF19} that asphericity would induce a significant error in spherical-collapse--based estimates of the turnaround radius; in contrast, they are in agreement with the findings of \citet{SouravTheodore2019}, that asphericity has a very minor effect on the turnaround radius. 

\subsection{Is there a dependence of the turnaround radius on the presence of massive neighbors?}\label{tides}

We have shown that asphericity is not the dominant factor driving deviations of the turnaround radius of a lower-mass simulated structure  ($\lesssim 10^{15} {\rm M_\odot}$ within $R_{ta}$) from the predictions of spherical collapse for a halo of identical mass. We have not however yet identified the culprit of the deviations in this mass range.

A hint comes from the average direction of the deviations. As seen in Fig.~\ref{Fig. densities}, the overall trend is towards higher values of $\rho_{ta}$. In the analytical treatment of \citet{SouravTheodore2019}, an increased value of the spherically-averaged turnaround density (decreased value of the spherically-averaged turnaround radius) is found in aspherical halos when the asphericity is driven by some effect opposing the gravitational attraction of the central potential, such as rotation. In contrast, a deviation from spherical symmetry itself, not driven by a gravity-opposing mechanism, produces to first order a zero net effect on the spherically-averaged turnaround radius. It is thus reasonable to conclude that we are looking for an effect which: (a) opposes the gravitational attraction of the central mass, and (b) is more likely to affect smaller halos.

\begin{figure}[!t]
\includegraphics[width=1.05\columnwidth]{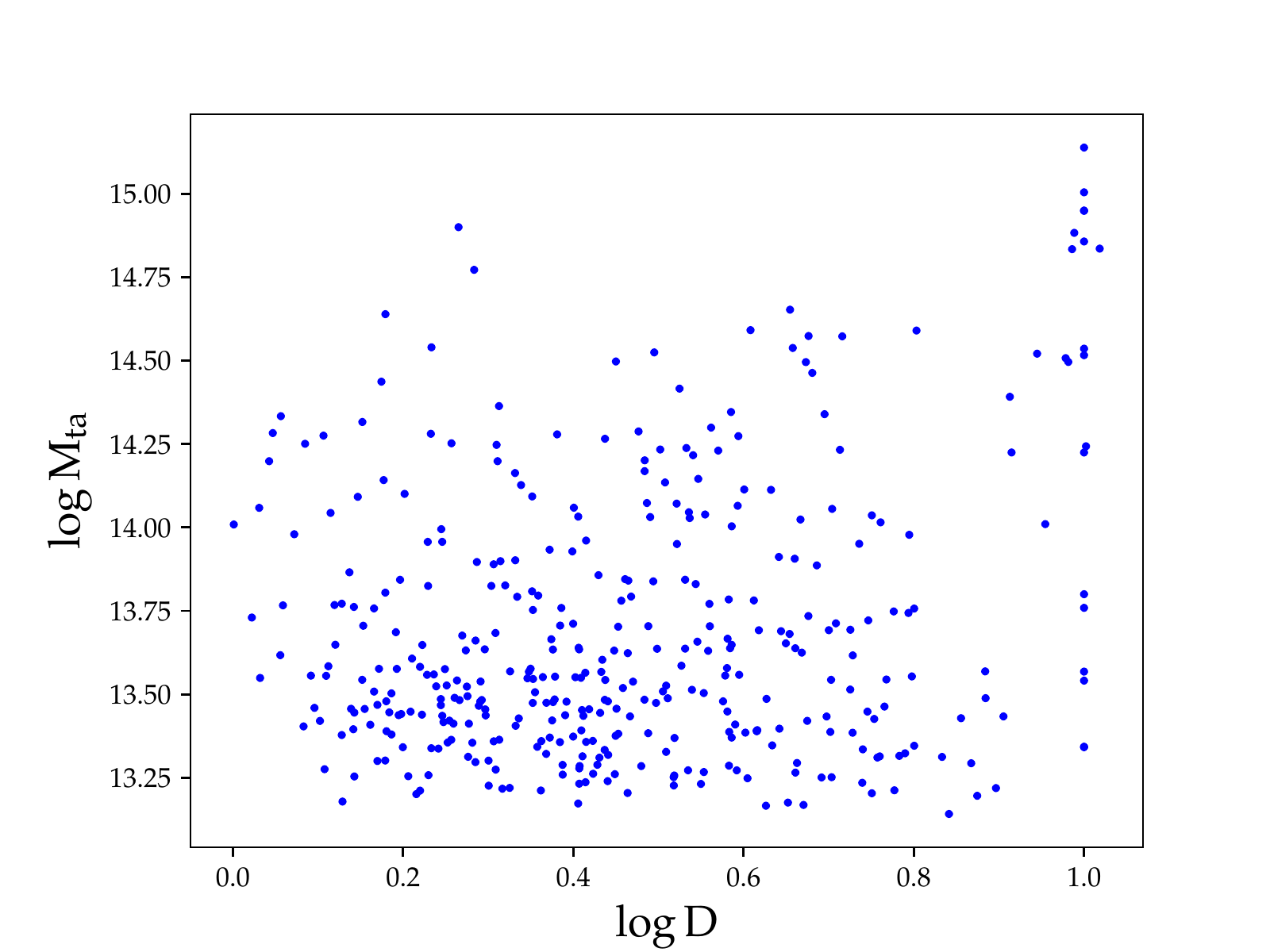}
\caption{Halo turnaround mass plotted against the tidal parameter $D$, for Illustris-TNG halos. The two are not significantly correlated (Spearman correlation coefficient $0.086754$, with a p-value of $9\%$). }
\label{Fig. 9.}
\end{figure}

The tidal effect of massive neighbors could play that role. It has a net effect opposing the gravity of the structure, and it is more likely to operate on small halos, as larger halos are too rare to be found nearby a neighbor of comparable or higher mass. To test this hypothesis, we select the neighbor with the dominant tidal effect, using an environmental parameter similar to ${D_{N,f}}$ of \citet{HaasEtal2012} (distance to the Nth nearest neighbor of a halo with $M_{ta}$ at least $f$ times the $M_{ta}$ of the halo, normalized to $R_{ta}$ of the neighbor). The tidal force due to the Nth nearest neighbor scales as $D_{N,f}^{-3}$. For simplicity, we select the neighbor within $10\times R_{ta,{\rm halo}}$ from the center of the halo with the mimimum value of ${D_{N,1}}$ as the one that will have the dominant tidal effect on the halo, and use that neighbor's $D-$value as a proxy for tidal effects suffered by the central halo. Our ``tidal parameter'' $D$ is therefore defined as
\begin{equation}
  D = \min \left. \left\{
\frac{r_{\rm neighbor-halo}}{R_{ta,{\rm neighbor}}}
\right|_{M_{ta, {\rm neighbor}}\geq M_{ta, {\rm halo}} \,\, {\rm and}  \,\, r_{\rm neighbor-halo} \leq 10\times R_{ta,{\rm halo}}}
\right\}\,. 
\end{equation}
An advantage of $D$ as a tidal parameter is that it is very weakly correlated, if at all, with halo mass \cite{HaasEtal2012}. In this way, any correlation identified cannot be attributed to a dependence on a common variable (mass). We verify this in our sample in Fig.\ref{Fig. 9.}, where we plot the turnaround mass of each halo against D, and show that they are not significantly correlated. 
\begin{figure}[!t]
\includegraphics[width=1.05\columnwidth]{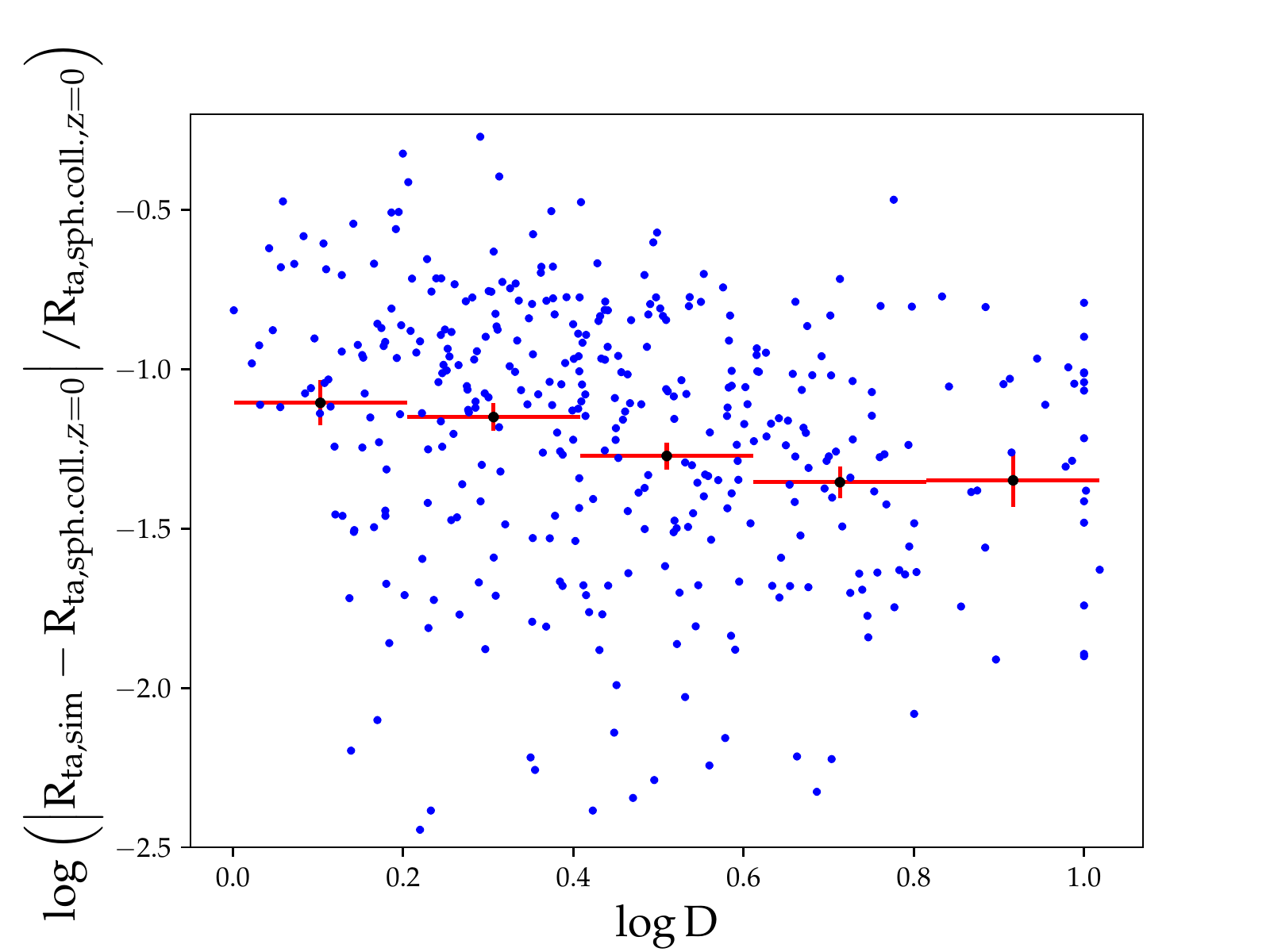}
\caption{Fractional deviation of $R_{ta}$ from the value predicted by spherical collapse for a structure of identical enclosed mass at $z=0$, plotted against the tidal paramer $D$ (see text). Lower values of D correspond to a stronger tidal force. A moderately strong (Spearman coefficient $-0.26$) but very significant (p-value $1.3\times 10^{-7}$) correlation can be seen.  The red datapoints correspond to the mean and standard error of the mean for the depicted bins in $\log D$. }
\label{Fig. 7.}
\end{figure}

\begin{figure}[!t]
\includegraphics[width=1.05\columnwidth]{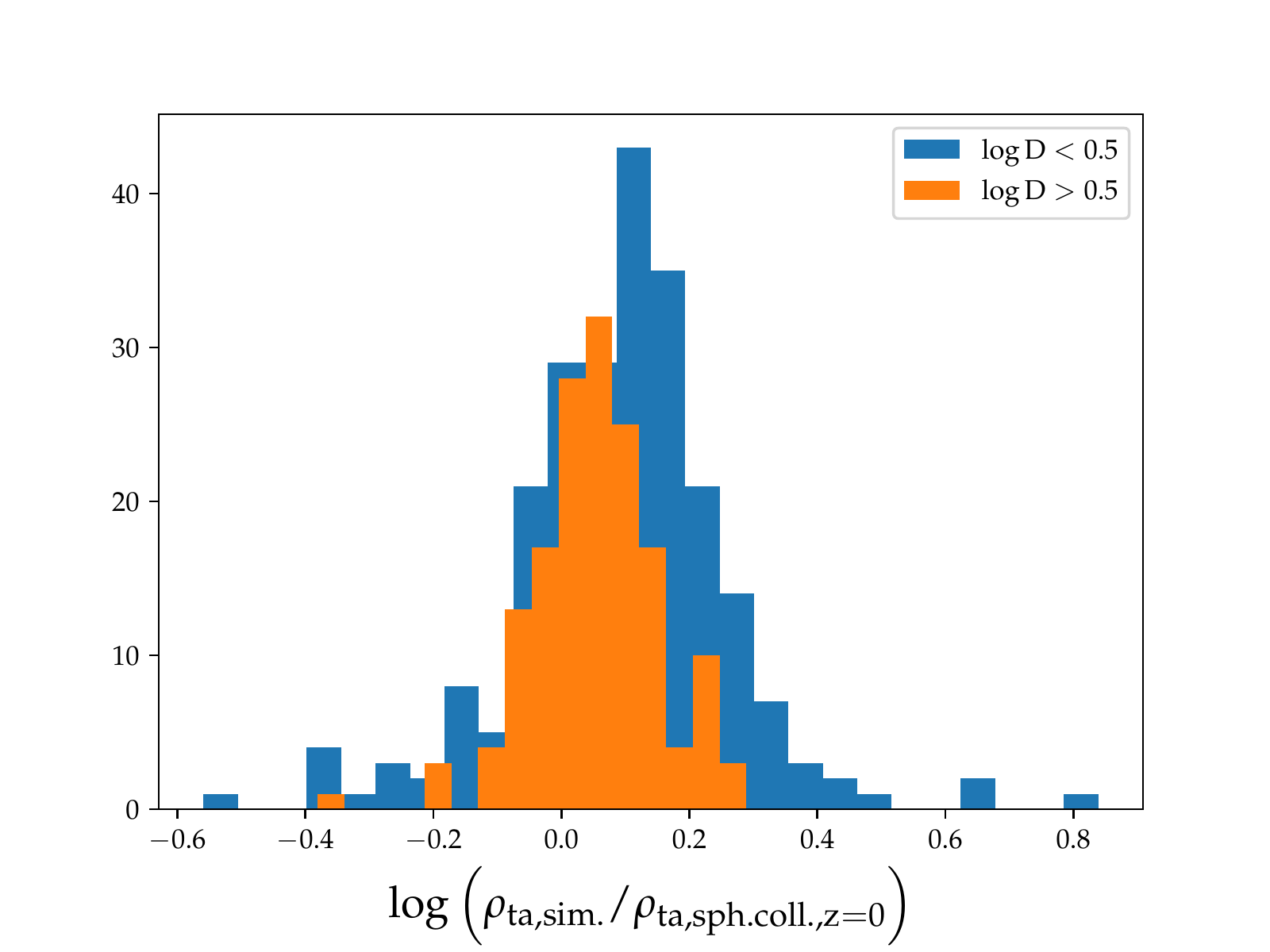}
\caption{Histogram of the logarithmic difference between $\rho_{ta}$ measured in Illustris-TNG halos and the predictions of spherical collapse. Blue histogram: halos with strong tidal effects from their environment ($\log D<0.5$). Orange histogram: halos with smaller tidal effects ($\log D \geq 0.5$). The medians of the two distributions are clearly and significantly shifted (blue: median  of $0.10$; orange: median of $0.05$. }
\label{Fig. 10.}
\end{figure}

In Fig.~\ref{Fig. 7.} we plot the fractional deviation of $R_{ta}$ from the spherical-collapse prediction against $D$, as a scatter plot for all halos and in bins of $ \log D$. The pileup of 10 halos at $D=1$ is because for the ten largest (and rarest) halos in Illustris-TNG no halo of comparable mass was found within $10\times R_{ta}$. Although we include these halos in the plot assigning to them a (lower-limit) value of $D=1$, we do not include them in the correlation analysis. A significant, moderately strong correlation is indeed found, with a Spearman correlation coefficient of $-0.26$ (more tidally affected halos deviate more from spherical symmetry predictions), with a p-value of $1.3\times 10^{-7}$. Note that for $\log D>0.6$ the effect flattens off, likely because the tidal effects of the dominant neighbor are already very weak at that point.

To verify that tidal effects of neigbors are indeed the dominant factor that drives the median $\rho_{ta}$ to lower values in smaller halos we show in Fig.~\ref{Fig. 10.} a histogram of the logarithmic difference between the turnaround density measured in Illustris-TNG halos and the predictions of spherical collapse, splitting halos according to the value of their tidal parameter, $D$. Indeed, halos with $\log D <0.5$ (which experience stronger tidal forces, blue histogram) have a higher median ($0.1$) than halos with $\log D \geq 0.5$ (median of 0.05). The latter is closer to the behavior seen in the larger TDSS halos.

\section{Discussion}\label{section discussion}

We have focused our discussion on halos with $M_{200}\geq 10^{13}{\rm \, M_\odot}$ (galaxy groups and higher in mass) for two reasons. The first is practical: these are the lowest-mass objects for which an observational determination of the turnaround radius could even in principle be attempted. Objects of significantly lower mass do not include enough galaxies to allow an accurate enough mapping of the Hubble flow and peculiar motions around them to determine the turnaround scale (e.g., \citealp{LocalGroup,Fornax,Virgo,Leesix}).  The second is physical: most frequently halos with smaller masses tend to be found in crowded environments, and neighbors larger than themselves in close proximity are very likely. This is problematic, not only because of the tidal effects of these neighbors, but because their presence contaminates significantly the radial velocity profiles of the small halos, complicating the determination itself of the turnaround scale in many cases.

We have explored in detail the possible effect of realistic shapes and environments on the turnaround radius of simulated dark matter halos. Similar issues have recently been discussed by \citet{Lee16}, who, extending the methodology of \citet{FSWB14}, introduced  empirical fits  to peculiar velocity profiles as a means to probe the exterior of virialized structures. \citet{Lee16} investigated the behavior of peculiar velocity profiles in cosmological simulations in the case where entire halos rather than dark matter particles are used to determine the peculiar velocity profile, and applied it specifically to the case of determining the turnaround radius.  \citet{LY16} used a similar methodology to study environmental effects on peculiar velocity profiles and estimate the likelihood that such effects might be responsible for observational estimates of the turnaround radius that apparently violate the bound $(3GM/\Lambda c^2)^{1/3}$ for the maximum radius of any non-expanding structure of \citet{Pavlidou_Tomaras}. Our findings here are consistent with these conclusions.

More recently, \citet{GF19} proposed a path towards investigating analytically the effect of deviations from sphericity on the turnaround radius, and predicted that asphericities would significantly affect the turnaround size of structures. Our analysis does not confirm this expectation. The effect of asphericity on the turnaround radius is only dominant in the largest, most rare objects, and even then the effect is very weak. 

In contrast, \citet{SouravTheodore2019} have calculated analytically that the effect of asphericities on the turnaround should be very small. When taking a spherically-averaged turnaround radius, they have predicted that the effect of asphericities should be non-vanishing to first order only when the asphericity is due to some gravity-opposing effect, such as rotation. Our analysis is consistent with these findings.   

The turnaround radius is not the only kinematically-motivated boundary of halos that can be defined. Other such boundaries proposed include the static radius (corresponding to the static mass) of \citet{Cuesta}, and the splashback radius of \citet{DK2014}. These are distinctly different from the turnaround radius we have studied here.

The static mass boundary corresponds to the innermost radius in which mean radial velocity is equal to zero - in other words, the innermost edge of the accretion region, within which the velocities are approximately randomized. The static part of the halo as used by \citet{Cuesta} (and the corresponding mass and radius) is thus the kinematically-defined equivalent of the virialized part of a halo. In contrast, the turnaround radius is the outermost zero crossing of the radial velocity - the outermost edge of any accretion region (which, for the larger-mass halos on which we focus, is always present).

The splashback radius is another boundary that is meant to separate virialized from infalling material, and thus is also located at the inner edge of the accretion region.  The splashback radius is defined as the radius where particles reach the apocenter of their first orbit, and corresponds to the first caustic of spherically-symmetric accretion.  The location of the splashback radius in simulated halos is identified  through analyzing either the density field around a halo \citep{MKD2017}, or the distribution of individual-particle orbit apocenters \citep{D2017}. 

The turnaround radius is located at much larger distances from the halo center than both the static and slpashback boundaries, and always outside the infall region. The motivation for studying it is that at these outermost non-expanding scales, the effect of the cosmological constant or any alternative-gravity effects on individual structures (rather than the Universe as a whole) would be most pronounced (e.g., \citealp{Pavlidou_Tomaras}, \citealp{BDRST17}, \citealp{Leesix}, \citealp{LVAS18}, \citealp{NOF18}, \citealp{CDL19}, Pavlidou et al. 2019 in prep.)

\section{Conclusions} \label{section 5}
In this paper, we used cosmological N-body simulations to probe the turnaround radius of dark matter halos with realistic shape, and compare its properties with those predicted from the spherical collapse model.

We found that a single turnaround radius can indeed describe well simulated halos, even in the presence of significant asphericities in their mass distribution. The value of the turnaround radius as measured from radial velocity profiles is in good agreement with spherical collapse for structures corresponding to large galaxy groups and galaxy clusters ($M_{200}>10^{13} M_\odot$).

Moreover, we showed that deviations of the turnaround radius from the spherical collapse model are primarily driven by the tidal forces of large nearby neighbors. The effect of the deviation of halo shapes from spherical symmetry is much weaker, and only detectable in larger halos, where the presence of nearby neighbors of comparable or larger mass is highly unlikely. 

Our results indicate that the turnaround radius could indeed be used as an alternative boundary for studying the abundance of massive large scale structures, as it is clean from the subtleties of baryonic effects and as it can be easily identifiable via the matter-density profile of structures. Indeed, although in our analysis 
the boundary of "turnaround structures" was identified using radial velocity profiles alone, these structures were all found to feature a characteristic average density.

This property suggests that halo-finders specifically geared towards the analysis of "turnaround" structures can be developed. For concordance cosmology, the predicted (by spherical collapse) density contrast of a region enclosed by the turnaround radius of a structure with the matter density of the background Universe is $\sim 11$. In this context, then, $R_{ta}$ is equivalent to $R_{11}$ -- in a way analogous to defining the "virial" radius as $R_{200}$. However, $R_{11}$ has the added benefit, shown in this work, to actually correspond well to the outermost zero-crossing of the spherically-averaged radial velocity profile for structures with $M_{200}>10^{13} M_\odot.$

Conversely, if the turnaround radius and mass of the halo on turnaround scales can be independently determined through observations, then the value of the turnaround density as a function of redshift can be used to probe cosmology.

\begin{acknowledgements}
G.K. would like to thank A. Zezas for his constructive comments throughout this work and G. Mouloudakis and E. Garaldi for fruitful discussions. G. K. and K. T. acknowledge support from the European Research Council under the European Union’s Horizon 2020 research and innovation program, under grant agreement No 771282. E. N. as a MC Fellow is supported by a Marie Curie Action of the European Union (Grant agreement number 749073). K.K. acknowledges funding from the European Research Council under the European Union’s Seventh Framework Program (FP/2007-2013)/ERC Grant Agreement n. 617001. This project has received funding from the European Union's Horizon 2020 research and innovation program  under the Marie Sklodowska-Curie RISE action,  grant agreement No 691164 (ASTROSTAT). 
\end{acknowledgements}
\bibliographystyle{aa}
\bibliography{bibliography}

\end{document}